\documentclass[usenatbib]{mnras}
\usepackage{newtxtext,newtxmath}
\usepackage[T1]{fontenc}

\DeclareRobustCommand{\VAN}[3]{#2}
\let\VANthebibliography\thebibliography
\def\thebibliography{\DeclareRobustCommand{\VAN}[3]{##3}\VANthebibliography}
\usepackage{xcolor}

\usepackage{graphicx}	% Including figure files
\usepackage{tcolorbox}
\usepackage{amsmath}	% Advanced maths commands
\usepackage{mathtools}
\usepackage{hyperref}
\hypersetup{colorlinks=true,
            citecolor=blue,
            linkcolor=blue}
\usepackage{enumitem}
\usepackage[export]{adjustbox}

\usepackage{chngcntr}
\usepackage{newtxtext,newtxmath}

\newcommand{\msun}{\mathrm{M}_{\odot}}	% Solar mass
\newcommand{\Hillradius}{r_{\mathrm{H}}}
\newcommand{\mBH}{M_{\mathrm{BH}}}
\newcommand{\mSMBH}{M_{\mathrm{SMBH}}}
\newcommand{\Rin}{R_{\mathrm{in}}}
\newcommand{\Rout}{R_{\mathrm{out}}}
\newcommand{\Rone}{R_1}
\newcommand{\Rtwo}{R_2}
\newcommand{\mdisc}{M_{\mathrm{d},0}}

\newlist{steps}{enumerate}{1}% create new steps environment
\setlist[steps]{%              configure the steps environment
  label=Step \Roman*,
  leftmargin=*,
  align=left
}
\tcbuselibrary{skins}
\tcolorboxenvironment{steps}{% wrap the steps environment in a tcolorbox
  colframe=red!0!black
}

\defcitealias{Rowan2022}{Paper~I}

\begin{document}

%%%%%%%%%%%%%%%%%%% TITLE PAGE %%%%%%%%%%%%%%%%%%%

% Title of the paper, and the short title which is used in the headers.
% Keep the title short and informative.
\title[Scatterings in accretion discs]{Black Hole Binaries in AGN Accretion Discs II: Gas Effects on Black Hole Satellite Scatterings}

% The list of authors, and the short list which is used in the headers.
% If you need two or more lines of authors, add an extra line using \newauthor
\author[C. Rowan et al.]{
Connar Rowan$^{1}$\thanks{E-mail: connar.rowan@physics.ox.ac.uk}, Henry Whitehead$^{1}$, Tjarda Boekholt$^{1}$, Bence Kocsis$^{1,2}$ and  Zoltán Haiman$^{3,4}$
\\
$^{1}$Rudolf Peierls Centre for Theoretical Physics, Clarendon Laboratory, University of Oxford, Parks Road, Oxford, OX1 3PU, UK
\\
$^{2}$St Hugh's College, St Margaret's Rd, Oxford, OX2 6LE, UK
\\
$^{3}$Department of Astronomy, Columbia University, New York, NY 10027, USA\\
$^{4}$Department of physics, Columbia University, New York, NY 10027, USA\\
}
% These dates will be filled out by the publisher
%\date{Accepted XXX. Received YYY; in original form ZZZ}

% Enter the current year, for the copyright statements etc.
%\pubyear{2015}

%\label{firstpage}
\date{\today}

\maketitle

\begin{abstract}
    The black hole (BH) binaries in active galactic nuclei (AGN) are expected to form mainly through scattering encounters in the ambient gaseous medium. Recent simulations, including our own, have confirmed this formation pathway is highly efficient. We perform 3D smoothed particle hydrodynamics (SPH) simulations of BH scattering encounters in AGN disks. Using a range of impact parameters, we probe the necessary conditions for binary capture and how different orbital trajectories affect the dissipative effects from the gas. We identify a single range of impact parameters, typically of width $\sim0.86-1.59$ binary Hill radii depending on AGN disk density, that reliably leads to binary formation. The periapsis of the first encounter is the primary variable that determines the outcome of the initial scattering. We find an associated power-law between the energy dissipated and the periapsis depth to be $\Delta E\propto r^{-b}$ with $b=0.42\pm0.16$, where deeper encounters dissipate more energy. Excluding accretion physics does not significantly alter these results. We identify the region of parameter space in initial energy vs impact parameter where a scattering leads to binary formation. Based on our findings, we provide a ready-to-use analytic criterion that utilises these two pre-encounter parameters to determine the outcome of an encounter, with a reliability rate of >90\%. As the criterion is based directly on our simulations, it provides a reliable and highly physically motivated criterion for predicting binary scattering outcomes which can be used in population studies of BH binaries and mergers around AGN. 
\end{abstract}

% do not make up new ones.

%%%%%%%%%%%%%%%%%%%%%%%%%%%%%%%%%%%%%%%%%%%%%%%%%%

%%%%%%%%%%%%%%%%% BODY OF PAPER %%%%%%%%%%%%%%%%%%

\section{Introduction}
Since the advent of gravitational wave detections \citep[e.g.][]{LIGO2016,LIGO2019,LIGO2020a,LIGO2020b,LIGO2020e,LIGO2020c,LIGO2020d,abbott2022,abbott2022_fermi}, understanding what astrophysical system(s) allow black holes to come close enough to one another for gravitational waves to dissipate their mutual orbital energy within a Hubble time is still a largely open problem. Suggested astrophysical scenarios that allow for such extremely close encounters include 3-body and binary-binary scatterings in star clusters \citep[e.g.][]{Zwart2000,Mouri2002,Miller2002,Downing2010,Rodriguez+2015,Rodriguez+2016,Rodriguez+2018,Samsing2018,Zevin+2019,DiCarlo2020,Liu2021,Arca_Sedda+2021}, isolated stellar binary evolution \citep[e.g.][]{Lipunov1997,Belczynski2010,Belczynski2016,Dominik2012,Dominik2013,Dominik2015,Tagawa2018}, hierarchical triple \citep[e.g.][]{Wen2003,Antonini+2012,Antonini+2016,Antonini+2017,silsbee2017_triple,Hoang+2018} and quadruple interactions \citep{Fragione_Kocsis2019}, and BH-BH encounters in quiescent galactic nuclei \citep[e.g.][]{OLeary+2009} and in active galactic nuclei AGN discs \citep[e.g.][]{Stone2017,Corley2019,Tagawa2020,McKernan2020_AGNmontecarlo,Li2021,Secunda2021,Li2022,Li_and_Lai_2022_windtunnel_II,Li_Dempsey_Lai+2022,Li2022_retroflows,Li_2022_hot_discs,Li_and_lai_2022,DeLaurentiis2022,Rozner+2022,Rowan2022,ford2022_AGNrates}. In the AGN scenario, the high density of the gas in an AGN disc is expected to harden the binary via gravitational or accretion processes. GW captures in the AGN channel may be identified from future observational data statistically using the angular power spectrum \cite[e.g.][]{Cusin2018,Gayathri2023}, coincidence with electromagnetic surveys \citep{Tagawa2023_flare,Rodriguez2023}, cross correlating the GW signal with AGN positions \cite[e.g][]{Bartos2017_localisation}, or through the distribution of source parameters, i.e. mass distribution extending to high BH masses consistent with hierarchical mergers \citep{Tagawa+2021_hierarchical,Tagawa+2021_mass}, large spin perpendicular to the binary orbital axis but the parallel spin component distributed around zero \citep[e.g.][]{Tagawa2020_spin,Tagawa+2021_hierarchical} , and non-zero eccentricities at 10Hz \citep{Tagawa+2021_eccentricity,Samsing+2022}. Individual smoking gun identification may be possible with LIGO/VIRGO/KAGRA by identfying an astrophysical GW echo due to GW lensing by the host SMBH \citep[e.g.][]{Gondan2022}, or by LISA measuring an acceleration of the center of mass \citep[e.g.][]{Leary2016,Inayoshi2021}, or by electromagnetic observations of a coincident transient \cite[e.g.][]{Graham2017,abbott2022_radio,abbott2022_gamma,Tagawa2023_flare,Tagawa2023_merge_obs,Tagawa2023_obs_emission}. 

The formation and evolution of black hole binaries in gaseous environments is not fully understood.
Simulations that consider purely gas driven evolution of a single isolated black hole binary (e.g neglecting three-body scattering, GW hardening, and the additional torque of the central massive SMBH) result in different outcomes depending on the assumed initial conditions. In the 2D simulations of \cite{Tang2017}, \cite{Munoz2019}, \cite{Moody2019}, \cite{Tiede2020} and 3D simulations of \citet{Moody2019}, which all adopted pressure conditions for a thick disc with scale height $H$ to radius $R$ ratio $H/R =0.1$, a concensus that binaries should experience net positive torques emerges, leading to an outspiral. However, the opposite is true for cooler, thinner discs \citep{Tiede2020, Heath2020}. It was also suggested that viscosity plays a role in cases where this transition occurs between positive and negative torques with respect to $H/R$ \citep{Dittman2022}. The modelling of subgrid/resolution environments around the BHs was also shown to affect the simulation outcome in some cases. The binary evolution was found to depend on the sink rate \citep{Tang2017} and softening lengths \citep{Li2021}, although more recent work found torques to have converged with respect to most numerical choices, at least for circular binaries~\citep[e.g][]{Moody2019,Munoz2020,Duffell2020,Westernacher-Schneider2022}. 

In the last few years there has been a steadily growing number of hydrodynamical simulations of binaries embedded in a SMBH accretion disc. In the first such simulations of \citet{Baruteau2011} (which consider stellar binaries), gas dynamical friction was found to harden a pre-existing binary in a 2D gas disc. This occured regardless of any gap formation in the SMBH disc, where the binary is sufficiently massive to expel gas from its orbit around the SMBH \citep[e.g.][]{Goldreich1980} more efficiently than it can be replenished through the disc's viscosity. \citet{Li2021} corroborated the binary hardening results of \citet{Baruteau2011} in the cases where the binary system orbit is retrograde with respect to its orbit around the disc/SMBH, as positive torque sources near the BHs are damped in the retrograde configuration due to disruption of the inner regions of their circum-single mini discs (CSMDs) that gravitationally deposit energy into the binary. For prograde binaries, both papers find their orbits expand over time. \citet{Kaaz2021} considered a binary in a thick disc where the binary does not open a gap in the disc using 3D wind tunnel simulations and find binary hardening in all their models. Such rapid hardening is also found in the three paper series of wind tunnel simulations of \citet{Li2022,Li_and_Lai_2022_windtunnel_II,Li_and_Lai_windtunnel_III_2023}. The source of the discrepancy between the wind tunnel simulations isn't clear, but could be the choice of equation of state (EOS), where the former assume an isothermal EOS and the latter consider an adiabatic EOS. In another paper that models the global AGN disc, \cite{Li_2022_hot_discs} find an enhanced temperature profile in the BH minidiscs lead to more pronounced negative torques and binaries that would normaly expand under isothermal conditions instead shrink. Extending the dimensions of the problem to 3D also leads to increased negative torques \cite[e.g][]{Dempsey2022}, although that study found that this reverses when the separation becomes 10\% of their Hill radius.
Other orbital elements may affect the evolution of the binary separation, such as eccentricity \citep{Munoz2019,DOrazio_Duffell2021,Siwek2023} and mass ratio \citep{Duffell2020,Dempsey2021,Siwek2023_accretion_mass_ratio}. For a review of how orbital elements can affect binary evolution, see  \cite{Lai_Munoz2023}.

While the overall evolution of embedded binaries is still a largely unconsolidated problem, the formation of such binaries is an even less understood topic. Highly detailed 1D semi-analytical work by \citet{Tagawa2020} suggested that of order $\sim90\%$ of binaries in AGN discs will have formed within the lifetime of the AGN disc. therefore understanding the initial orbital elements of these in situ formed binaries is essential to understanding their later evolution. These binary formation events are expected to predominantly take place through what is known as a \textit{gas assisted} capture whereby upon the scattering of two BHs in the AGN disc, a complex interplay between the binary and the surrounding gas and their CSMDs leads to a net removal of the binaries' kinetic energy \citep[e.g][]{Goldreich2002}, such that they remain bound. This was investigated very recently in semi-analytical studies \citep[e.g.][]{DeLaurentiis2022,Rozner+2022}. These studies assumed orbital energy is dissipated by an \citet{Ostriker1999} like dynamical friction prescription between the binary BHs and the local gas. Finding higher AGN densities favour binary formation in BHs with smaller radial separations in the SMBH disc. The first full hydro simulations of this process \citep{Rowan2022,Li_Dempsey_Lai+2022} corroborated the overall analytical and semi-analytical expectation that scattering encounters in gas may lead to binary formation, even when their initial orbital energy is positive upon first entering each others' Hill sphere, but the details may differ from those of previous semianalitical models. In our previous paper (\citet{Rowan2022}, hereafter \citetalias{Rowan2022}), the subsequent evolution of the binaries was also directly simulated following their formation, showing that their evolution depends on the AGN disc mass and their orbital configuration (i.e prograde or retrograde). Inspiral was observed in some prograde cases and for retrograde binaries, eccentricities were excited to high enough values in short enough times for binary mergers to occur within timescales of only a few AGN orbits. Both \citetalias{Rowan2022} and \citet{Li_Dempsey_Lai+2022} found inpiraling binaries that form with very high initial eccentricities, possibly explaining the discrepancy with studies that used near circular initial binaries \cite[e.g.][]{Baruteau2011,Kaaz2021,Li_and_lai_2022,Li_2022_hot_discs}. \citet{Li2022} further corroborated this with a positive correlation between eccentricity and inspiral rate. 

Now that the viability of the gas assisted capture mechanism has been justified, the next natural question is to quantify the efficiency of this process based on pre-encounter parameters. Our previous gasless study \citep[][]{Boekholt_2022} identified three `islands' in the parameter space of the impact parameter (i.e. initial radial separation relative to the central SMBH) $p$ of black hole encounters that lead to more than one encounter, i.e. ``a Jacobi capture''. The dependence of the number of encounters (orbits) during the Jacobi capture on the impact parameter exhibits a fractal structure. Furthermore, the large statistics of the work allowed one to very finely sample the parameter space of the encounter to quantify the fraction of the parameter space $p$ that lead to temporary binary formation. Here, we examine if these temporary binaries formed without gas in \citet{Boekholt_2022} lead to permanent capture through gas hardening. We also studied how gas affects the fractal structure of the parameter space of initial conditions that allow for successful binary capture. In the first paper of this series \citetalias{Rowan2022} and in the work of \citet{Li_Dempsey_Lai+2022}, the gas capture process was validated using a full hydrodynamical approach for the first time. In \citetalias{Rowan2022} the formation and long term evolution of each binary was modelled self consistently in one continuous simulation, so the starting conditions of the binaries were directly set by their formation and not based on any assumptions. The work demonstrated that both prograde and retrograde binaries may form in varying AGN gas densities and identified a bimodality in the eccentricity evolution where prograde binaries circularised over time and retrograde binaries became more eccentric.  

In this paper we explore the regions of phase space that permit the formation of a binary from two initially isolated black holes embedded in a gaseous accretion disc of a third massive body (SMBH) and the role of gas in shaping this space  using a global 3D hydrodynamical simulation based on smoothed particle hydrodynamics (SPH). This work is the second in a series of papers investigating each element of the gas-induced binary merger mechanism in AGN. Our investigation builds on \citetalias{Rowan2022} by probing the binary capture process in far greater detail using a much finer sampling (114 simulations total) in impact parameter, providing great enough statistics to probe the stochasticity of the gas-assisted formation process and develop statistical correlations between pre-encounter parameters and important physical processes during the encounter. Such correlations can then be used to generate a vastly improved analytic prescription for semi-analytic works that discuss binary formation in accretion discs like that of \citet{Tagawa2020}. We present this paper alongside a companion paper \citet{Henry_inprep} which models the same scattering problem considered here but using a local shearing box approach and employing the grid code Athena++ instead of our SPH implementation. As we will discuss later in detail, the results of both papers are largely consistent with one another. Such similarities include the relationship betweeen the initial approach trajectory of the BHs and the level of dissipation and the dependence on gas density.
The numerical method and initial conditions are detailed in Sec. \ref{sec:CompMethods}, and the results are presented in Sec. \ref{sec:results}. We discuss our results in Sec. \ref{sec:discussion} and the caveats of the models in Sec. \ref{sec:caveats} before summarising and concluding in Sec. \ref{sec:conclusions}.     
\section{Computational Methods}
\label{sec:CompMethods}
\subsection{Hydrodynamics}
We consider a three-body system consisting of a central SMBH of mass $\mSMBH=4\times10^{6}\msun$ and two stellar BHs with masses $\mBH=M_{1}=M_{2}=25\msun$ embedded in a gas disc. The BHs are inserted symmetrically about a radial encounter distance from the SMBH, $R_{\rm{mid}} = 0.0075\,\mathrm{pc}$. The gas disc is resolved about the BHs as an annulus of radial width $p_{\rm{disc}} = 20\Hillradius$ where
\begin{equation}
    \centering
    \Hillradius = R\bigg(\frac{\mBH}{3\mSMBH}\bigg)^{1/3},
    \label{eq:Hill}
\end{equation}
is the Hill radius of one of the satellite BHs, from an inner radius, $R_{\rm in}$, to outer radius, $R_{\rm out}$. Based on our annulus width this gives $R_{\rm in}=R_{\rm{mid}}-10\Hillradius$ and $R_{\rm out}= R_{\rm{mid}}+10\Hillradius$.
The hydrodynamics of our problem are simulated using the 3D smooth particle SPH code PHANTOM \cite[see][]{Price2018} utilising 25 million particles, ignoring radiative and magnetic effects which for the distances we consider from the SMBH are less relevant \citep{Jiang2019,Davis2020}. Such effects may be important for modelling the CSMDs of the satellite BHs, however due to the computational expense of incorporating them at our resolution, we also ignore them here. The AGN disc, where we will simulate the encounter of two BHs, is consistent with its implementation in our previous work in \citetalias{Rowan2022}, though we summarise its features here for completeness. The disc is represented by a thin Shakura-Sunyaev constant alpha disc (\citealt{Shakura1973}) with a viscosity parameter $\alpha_{\rm{SS}}=0.1$ and scale height to radius ratio of  $H/\Rin=0.005$. This value corresponds to an accretion rate at this radius of 10\% Eddington \citep[see][]{Goodman2004}.

The enclosed disc mass as a function of radius is assumed to follow the power-law of that derived in \citet{Goodman2004} which is defined with four additional free parameters
\begin{multline} 
M_{\rm d,0}(<R)\approx 4.82\times10^{5}\alpha_{SS}^{-\frac{4}{5}}\hat{\kappa}^{-\frac{1}{5}}\xi^{-\frac{4}{5}} \\ 
\times\bigg(\frac{L_{E}}{\epsilon}\bigg)^{\frac{3}{5}}\bigg(\frac{M_{\rm SMBH}}{10^{8}\msun}\bigg)^\frac{11}{5}\bigg(\frac{R}{1000r_{s}}\bigg)^{\frac{7}{5}}\msun.
\label{eq:discmass}
\end{multline}
Here $l_{E}$ is the disc luminosity relative
to the Eddington limit, $\xi$ is the mean molecular mass, $\hat{\kappa}$ is the opacity and $\epsilon$ is the radiative efficiency. The disc has a sound speed $c_{\rm s}$ profile given by 
\begin{align}
    c_{\rm s}(R) &= c_{s,\rm in}\left(\frac{R}{\Rin}\right)^{q}\,,\label{eq:cs_prof}\\
    c_{\rm s,\rm in} &= \left(\frac{H}{R}\right)_{\Rin}
    \sqrt{\frac{GM_{\rm SMBH}}{\Rin}}
    \,,
    \label{eq:cs0}
\end{align}
where the $c_{\rm s,in}$ is the sound speed at $R_{\rm in}$ and q is the power-law exponent. The surface density $\Sigma$ profile is similarly represented as 
\begin{equation}
    \centering
    \Sigma(R) = \Sigma_{0}\bigg(\frac{R}{\Rin}\bigg)^{-f}\,,
    \label{eq:rho_R}
\end{equation}
with its own radial dependence $f$. The value of $\Sigma_{0}$ calculated through normalisation of \eqref{eq:rho_R} such that
\begin{equation}
    \centering
    2\pi\int_{\Rin}^{\Rout}  \Sigma(R)R dR = \mdisc(\Rout) - \mdisc(\Rin)\, .
    \label{eq:sig_norm}
\end{equation}
That is to say the radial integration of the surface density across the annulus matches the expected mass of the annulus based on \eqref{eq:discmass}. This gives a 3D density distribution of 
\begin{equation}
    \centering
    \rho(R,z) = \frac{\Sigma(R)}{\sqrt{2\pi}H}\exp \bigg(\frac{-z^{2}}{2H^{2}}\bigg).
    \label{eq:rho_Rz}
\end{equation}
\subsection{Initial Conditions}
In a similar manner to our previous work of the gasless case in \citet{Boekholt_2022} we 
%want to observe 
examine how binary formation varies with the approach of the binary, namely the impact parameter $p$, closest approach $r_{\min,1}$ and relative two-body energy in the presence of gas with disk mass $\mdisc$ given by Eq.~\eqref{eq:discmass}. To do this we initialise the two BH satellites on circular orbits around the SMBH with varying initial radial separations which we will denote as their impact parameter $p=\Rtwo-\Rone$ where $\Rone$ and $\Rtwo$ are the respective radial positions of the inner and outer satellite in the SMBH disc. The impact parameter $p$ is sampled using 39 evenly spaced points over the range $[1.75,4.25]\Hillradius$. This is an $\sim 8\times$ increase in the number of simulations compared to \citetalias{Rowan2022} which only considered 5 impact parameters per set of initial parameters. In the previous paper, this limitation prohibited any statistical analysis of the effect of varying impact parameters. Now, using far more models, we can investigate this region of the parameter space for BH binary formation. Recent literature indicates that binaries can dissipate their relative energy more efficiently in denser AGN discs \citep[e.g.][]{Li_Dempsey_Lai+2022,DeLaurentiis2022,Rowan2022}. Hence, we also probe more rigorously how the capture window is tied to the density of the surrounding accretion disc by performing another sample of 45 simulations using an AGN disc with a mass of $3\mdisc$. We also compare to a gasless sample with the same initial conditions for reference \citep[see][for a detailed study of the gasless case]{Boekholt_2022}. This gives the mass of gas contained within an annulus of width $\Hillradius$ of $0\msun, 110\msun$ and $330\msun$ for the gasless, fiducial and $3\mdisc$ models respectively, indicating the high amount of gas present in our simulations.

\begin{table*}
    \centering
    
    \begin{tabular}{ccccccccccccccc}
    \hline\hline
$p$ & $\Delta \phi$
& $\dfrac{M_{\rm SMBH}}{\msun}$ 
& $\dfrac{M_{\rm BH}}{\msun}$ 
& $\dfrac{M_{\rm d}}{10^{-3}M_{\rm SMBH}}$ 
&  $\dfrac{R_{\rm mid}}{\rm mpc}$ & $\dfrac{p_{\rm disk}}{\Hillradius}$ 
&  $f$ & $q$ & $\dfrac{H}{R_{\rm in}}$ & $\alpha_{\rm SS}$  %\\ \hline
 %\\ \hline
&    $L_{\rm E}$ & $\epsilon$ & $\xi$ & $\hat{\kappa}$ 
\\ \hline 
    1.75--3
    & $20^{\circ}$ 
    & $4\times10^{6}$
    & 25 
    & $\{0,1.6,4.8\}$
    & 7.5 & 20 
    & 0.6 & 0.45 & 0.005 & 0.1 
    &  0.1 & 0.1 & 0.6 & 1.0 
    \\ \hline
    \end{tabular}
    \caption{Fiducial model parameters. Here $(p, \Delta \phi)$ are the initial radial offset and orbital phase between the two stellar BHs in their initially Keplerian orbits around the SMBH, $\Hillradius$ is the Hill radius, $(M_{\rm SMBH}, M_{\rm BH}, M_{\rm d})$ are respectively the SMBH mass, the individual stellar BH masses, and the total enclosed gaseous disk mass (Eq.~\ref{eq:discmass}), hence the gas mass per $\Hillradius$ radial width is $\frac75 (\Hillradius/R_{\rm mid})M_{\rm d}=0.026 M_{\rm d}=\{0,\,110,\,330\}\msun$, 
    $(R_{\rm mid},p_{\rm disk})$ are the mean radius and the width of the simulated gaseous annulus, $q$ and $f$ set the radial dependence for the sound speed and surface density across the annulus (Eqs.~\ref{eq:cs0}--\ref{eq:rho_R}), $H$ is the scaleheight ($H=0.4\,\Hillradius$) which also sets the pressure and temperature in the disk via Eq.~\eqref{eq:cs0}, $L_{\rm E}$ is the Eddington ratio, $\epsilon$ is the radiative efficiency, $\xi$ is the mean molecular mass, $\hat{\kappa}$ is the opacity relative to the electron scattering opacity ($0.4\,{\rm cm}^{2}{\rm g}^{-1}$).
}
    \label{tab:initial_conditions}
\end{table*}

As we are now including gas in this experimental setup, there are several additional complexities that can affect the interpretation of the results compared with the gasless case. First, BHs form circum-single mini disks (CSMDs) before their encounter. The discs enhance the gravitational attraction between the satellite BHs since their masses can be approximately added to their host BH to leading (monopole) order in a multipole expansion. Additionally, the masses of the BHs can change via accretion in the lead up to the encounter which again can alter the amount each satellite perturbs the other prior to the encounter itself. To have a more similar set of BH and CSMD masses at their encounter we scale the initial azimuthal separation $\Delta \phi$ of each object, depending on their impact parameter $p$, such that their approach time is approximately the same. Specifically, $\Delta \phi$ is scaled so all simulations have the same approach time as that with $p=2.5\Hillradius$ and $\Delta \phi=20^{\circ}$, i.e
\begin{equation}
    \centering
    \Delta\phi(\Rone,\Rtwo)= \frac{\sqrt{\frac{\mSMBH}{\Rone}}-\sqrt{\frac{\mSMBH}{\Rtwo}}}{\sqrt{\frac{\mSMBH}{R_{\rm mid}-1.25\Hillradius}}-\sqrt{\frac{\mSMBH}{R_{\rm mid}+1.25\Hillradius}}}\times20^{\circ}\,.
    \label{eq:phi_i}
\end{equation}
This ensures that the amount of accretion and growth of the CSMDs is approximately equal for each simulation with different $p$. We summarise our simulation parameters and initial conditions in Table \ref{tab:initial_conditions}.
\section{RESULTS}
\label{sec:results}
In this section, we present and discuss the results of our simulation suites. Starting with an overview of the shape and size of the formation parameter space in Sec. \ref{sec:cross_sect}, how the energy dissipation of the encounter is influenced by the closest approach is discussed in Sec. \ref{sec:minsep}. The time dependence of each mechanism that can alter the orbital energy of the binary during the encounter is described in Sec. \ref{sec:mechanisms}. Accretionless encounters are considered in Sec. \ref{sec:Accretionless encounters} and the effects of an increased AGN disc density are presented in Sec. \ref{sec:diff_density}.
\subsection{The capture cross section with gas driven dissipation}
\label{sec:cross_sect}
\begin{figure}
    \centering
    \includegraphics[width=9cm]{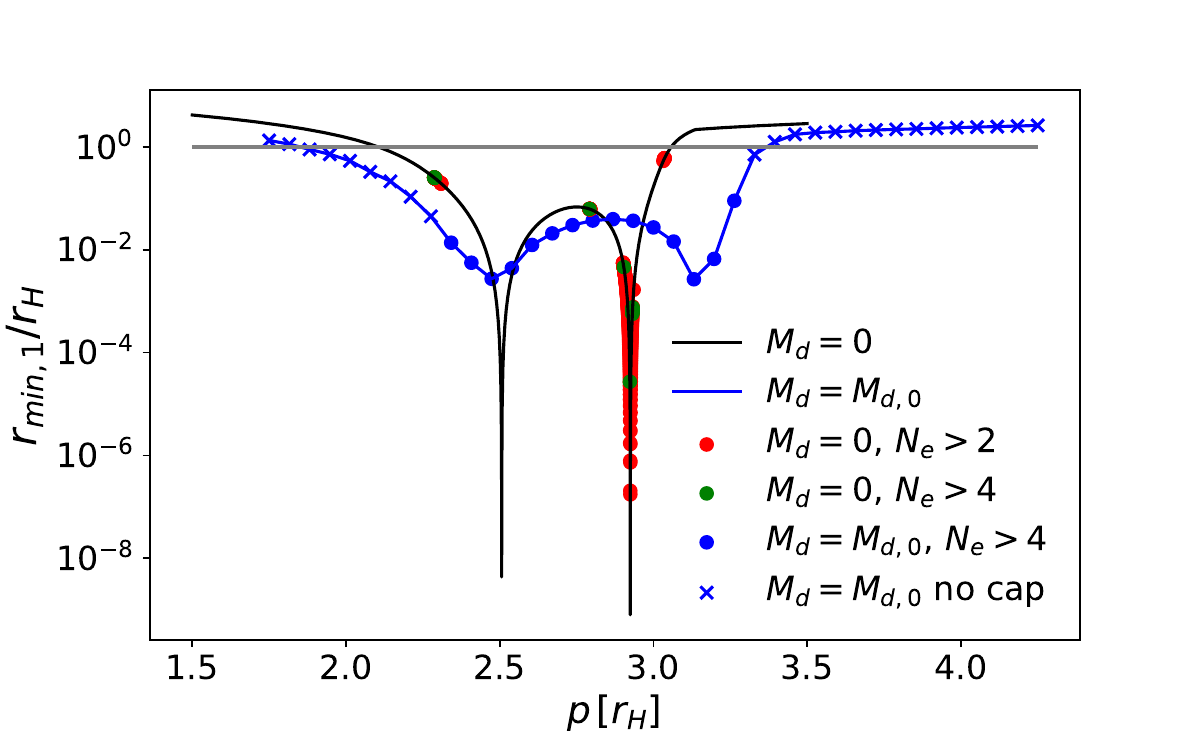}
    \caption{Minimum separation for our fiducial run (blue) compared to the gasless case (black) for the first approach as a function of the ``impact parameter'', i.e. the initial radial separation relative to the SMBH, denoted by $p$. The horizontal line indicates the Hill radius of one of the objects for comparison. The gas clearly broadens the range of impact parameters of close encounters. In the gasless $M_{\rm d}=0$ simulation suite, binaries with a number of encounters of $N_{e}\geq2$ and $N_{e}\geq4$ are shown in red and blue respectively. For the our hydrodynamic simulation, all binaries with two encounters also performed four and remain bound.
    }
    \label{fig:minsep}
\end{figure}

The periapsis distance of each first encounter is shown in Figure \ref{fig:minsep}. These are compared directly to resimulated gasless encounters using the exact same initial conditions for the BHs. As the computational cost without gas is orders of magnitude lower, a much finer resolution of 16k (i.e. $2^{14}$) models with impact parameters in the range $b=[1.5,3.5]$ are used. It is immediately obvious that gas leads to a much larger window in impact parameters for encounters to penetrate less than a Hill radius. Moreover, there is only one wide valley of captures in the space of $p$, with the minimum separations varying gradually across $p$, unlike in the gasless case where there are two separated capture windows and sharp changes in the minimum separation for very particular initial conditions (as identified in \citealt{Boekholt_2022}). Note that the deep troughs observed in the gasless case are possibly missed by the relatively coarse sampling in impact parameters for the simulations with gas. Additionally, gas leads to binary formation at depths far shallower than in the gasless simulations, i.e at larger separation at the first closest approach, $r_{\min,1}$.
In order to quantify directly the size of the parameter space in $p$ where multiple encounters may happen, we define the initial one-dimensional capture \textit{cross section} $\lambda$ as the total "length" in the 1D parameter space of the impact parameter $p$ that allows for at least \textit{two} encounters within the binary Hill sphere. Even though many gasless and gaseous encounters later decouple after this point, we examine the dissipation required for a binary to initially be bound, not whether the system is stable to the 3-body interaction with the SMBH in the long term. More generally, the cross section for encounters to have exactly $n$ encounters within the Hill sphere is defined in \citet{Boekholt_2022} as
\begin{equation}
    \lambda\left( N_e = n \right) = \sum_i \Delta \lambda_i \delta_{n, N_{e,i}},
\end{equation}
where $N_e$ is the number of encounters of a simulation, $i$ represents the simulation number, $\delta$ is the Kronecker delta which vanishes unless simulations have $n$ total encounters and $\Delta \lambda_i = \frac{1}{2}\left(p_{\rm{i+1}} - p_{\rm{i}}\right)$ is the length of the impact parameter interval between the discrete samples around the simulation $i$. Here we are interested in the cross section for \textit{all} models which have $N_e \geq 2$. So we replace the Kronecker delta with an indicator function $I_A(x)$
\begin{equation}
I_{A}(x)=\left\{
\begin{array}{ll}
      1 & \mathrm{if~} x \in A \\
      0 & \mathrm{otherwise} 
\end{array} 
\right., 
\end{equation}
that vanishes if case $x$ is not part of set $A$. For our problem, this becomes
\begin{equation}
I_{[N_e \geq 2]}(N_{e,i})=\left\{
\begin{array}{ll}
      1 & \mathrm{if~} N_{e,i} \geq 2 \\
      0 & \mathrm{otherwise}  \\
\end{array} 
\right..
\end{equation}
All together this gives explicitly the initial capture cross section $\lambda$
\begin{equation}
    \lambda\left( N_e \geq 2 \right) = \sum_i \Delta \lambda_i I_{[N_e \geq 2]}(N_{e,i}),
    \label{eq:cross_sect}
\end{equation}
with an associated error $\delta\lambda$ of
\begin{equation}
    \delta\lambda\left( N_e \geq 2 \right) = \sqrt{\frac{1}{2}\sum_i \bigg(\Delta \lambda_i I_{[N_e \geq 2]}(N_{e,i})\bigg)^{2}}.
\end{equation}
We compute this quantity based on all $2^{14}$ gasless and all 39 of our fiducial gaseous simulations. We get a cross section for two encounters of $\lambda(N_{e}\geq2)=(0.0407\pm0.0011)\,\Hillradius$ for our gasless simulations and $0.98\pm0.13\Hillradius$ when gas is included. The cross section for four encounters then drops to $(\lambda(N_{e}\geq4)=(0.00219\pm0.00025)\,\Hillradius$ for the gasless case but remains identical for our simulations with gas. The drop in the gasless case follows as there is no dissipation mechanism to remove energy once the binary is formed other than through its interaction with the SMBH which is entirely chaotic, thus they all inevitably decouple after usually only a short number of orbits after being disrupted by the SMBH, see \citet{Boekholt_2022} for details. The fact $\lambda(N_{e}\geq2)=\lambda(N_{e}\geq4)$ when gas is included indicates that gas can dissipate their orbital energy rapidly enough so that all binaries that undergo two encounters harden sufficiently to prevent their disruption from the SMBH. Furthermore, based on these values we conclude that for our initial conditions, there is a $\sim25$ times larger cross section to have multiple orbits when gas is included based on $\lambda(N_{e}\geq2)$ with and without gas, indicating that binaries are at least temporarily captured far more efficiently in the presence of gas. Whether these binaries are permanent is examined below. Directly comparing the $\lambda(N_{e}\geq2)$ and $\lambda(N_{e}\geq4)$ cross sections of our gasless simulations to those of \citet{Boekholt_2022} ($\lambda(N_{e}\geq2)=0.12r_{\mathrm{H}}$, $\lambda(N_{e}\geq4)=0.0051r_{\mathrm{H}}$) we find slightly lower cross sections while still within the same order of magnitude for both values, though we reiterate that that paper used a far larger initial azimuthal separation so the gravitational focusing will of course be different across those simulations compared to the simulations in this investigation. The slope of $d \log_{10}(\lambda)/dN_e$ was shown in \citet{Boekholt_2022} to be constant at $\sim0.68$. We calculate this slope to be $\sim 0.64$ , in good agreement.
\subsection{Dissipation as a function of minimum separation}
\label{sec:minsep}
Though the encounters lead to largely chaotic gas flows, we develop ready-to-use semi-analytical prescriptions based on our high-resolution simulations to improve the accuracy of purely semi-analytical studies such as by \cite{Tagawa2020}, \cite{DeLaurentiis2022} and \cite{Rozner+2022}. In this work, we define explicitly the \textit{first encounter} as the period between the BHs entering within 2$\Hillradius$ separation and either the first apoapsis or upon reaching $1\Hillradius$ separation after the first encounter where the choice of end point depends on whether the binary remains bound or not, respectively. If the binary becomes bound (even if temporarily) we define the \textit{second encounter} as the period between first apoapsis and second apoapsis or the time of exiting the Hill sphere, and so on for higher numbers of encounters. The two-body energy of the binary is given by 
\begin{equation}
    \centering
    E_{\rm bin}=\frac{1}{2}\mu \|\boldsymbol{v}_1-\boldsymbol{v}_2\|^2 - \frac{GM_{\rm{bin}}\mu}{\|\boldsymbol{r}_1-\boldsymbol{r}_2\|},
    \label{eq:two_body_energy}
\end{equation}
where $M_{\rm{bin}}=M_{1}+M_{2}$ is the total mass of the binary, $\mu = M_{1}M_{2}/M_{\rm{bin}}$ is the reduced mass, $\boldsymbol{r}_i$, $\boldsymbol{v}_i$ are the positions and velocities of satellite BH $i=(1,2)$ and $G$ is the gravitational constant. 

\begin{figure}
    \centering
    \includegraphics[width=9cm]{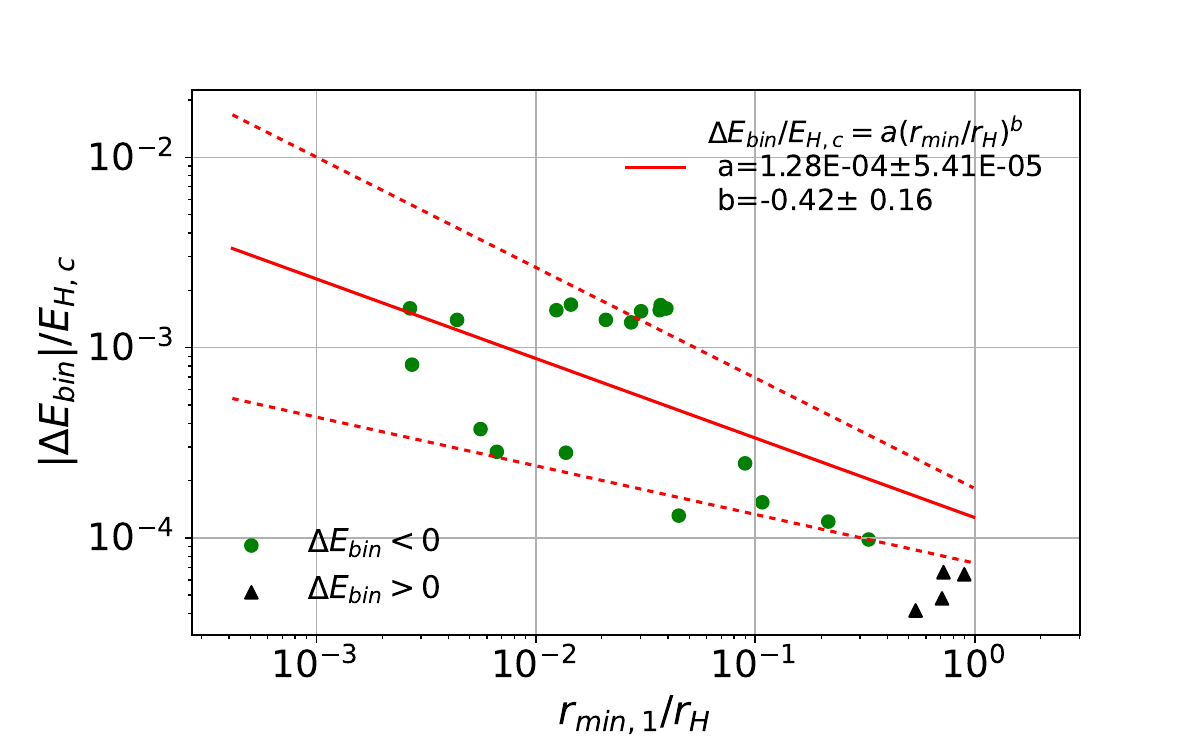}
    \caption{The change in the binary energy during the first encounter, $\Delta E_{\rm bin}$ normalised to $E_{\rm H,c}$, as a function of the first periapsis passage $r_{\min,1}$. The sign of $\Delta E_{\rm bin}$ is indicated by the symbol type: circular green points indicate where energy is \text{removed} from the binary while the black triangular points represent models where energy is \textit{added}. Only binaries that pass within $\Hillradius$ are shown since all encounters outside the Hill sphere result only in flyby encounters. The red  solid and dashed lines show respectively the best-fit power-law relation to the dotted data points and the $1\sigma$ errors on the slope.}
    \label{fig:dE_first_enc_fit}
\end{figure}

Figure \ref{fig:dE_first_enc_fit} shows $\Delta E_{\rm bin}$, the change of the binary energy during the first encounter as a function of closest approach in our fiducial simulation setup expressed in the natural units of $E_{\mathrm{H},0}$ defined as
\begin{equation}
    \centering
    E_{\rm H,c} = \frac{GM_{\rm{bin}}\mu}{2\Hillradius}\,,
    \label{eq:E_H}
\end{equation}
i.e. the orbital energy of a binary with a semimajor axis equal to the Hill sphere of one of the BHs, where the sign of $\Delta E_{\rm bin}$ is denoted by different point types as indicated in the caption. 
%BK: Note that I removed mass-specific, since this has units of energy. }
Also plotted is a best-fit line for the analytic power-law profile as
\begin{equation}
    \centering
    \Delta E_{\rm{bin}} = a\bigg(\frac{r_{\min, 1}}{\Hillradius}\bigg)^{b}E_{\rm H}\,.
    \label{eq:powerlaw}
\end{equation}
where $r_{\min,1}$ is the minimum separation of the first encounter, $a$ and $b$ are dimensionless fitting parameters.

\begin{figure}
    \centering
    \includegraphics[width=9cm]{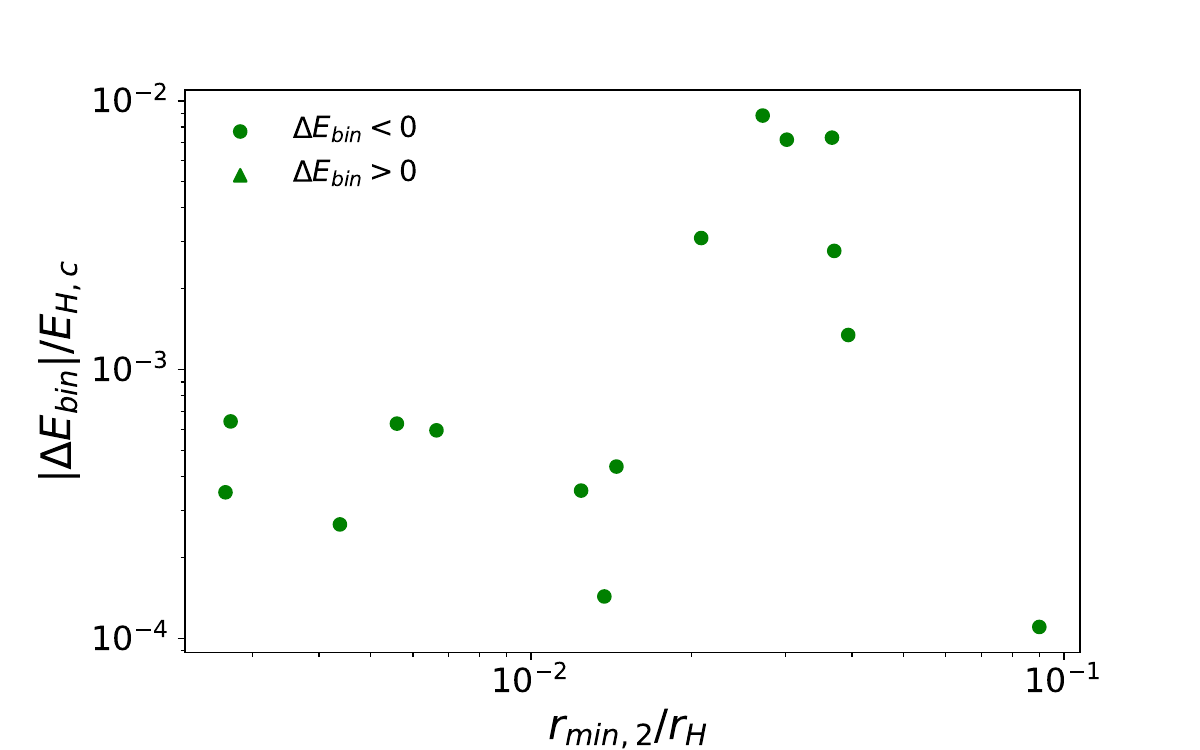}
    \caption{Similar to Figure~\ref{fig:dE_first_enc_fit} but showing the change in the binary energy during the second encounter, normalised to $E_{\rm {H,c}}$ as a function of the closest approach during the second encounter $r_{\min,2}$. Note that there are now naturally fewer data points as many models decouple after only their first encounter.}
    \label{fig:dE_second_enc}
\end{figure}

Figure \ref{fig:dE_first_enc_fit} indicates an anticorrelation trend between the depth of the first encounter and the amount of energy dissipated, i.e deeper encounters tend to dissipate more energy.\footnote{The existence of this dissipation-periapsis relation was hinted in \citetalias{Rowan2022}, where dissipation was found to be more efficient when binaries directly intersect each others' CSMDs although not quantified until now.} There is a large scatter where the energy dissipated in a small bin of closest separations can vary by up to two orders of magnitude for small $r_{\min,1}$. This scatter may be interpreted as an error in the predicted power-law exponent $b$ as shown. For our parameters, the binary objects intersect each others' CSMDs at $\Delta r \sim 0.05r_{\mathrm{H}}$. Thus since the trend continues outside of this range, the coupling between the dissipation and periapsis changes continuously with $\Delta r/\Hillradius$, rather than akin to a step function at the intersection of the BH CSMDs as previously thought in \citetalias{Rowan2022}.
There is no correlation for the second and later encounters (Figure \ref{fig:dE_second_enc}), as the gas morphology which determines the amount of energy dissipation becomes chaotic. 
The first encounter is naturally the most important for discussing energy dissipation as if the binary energy is still too high after the first encounter then no further encounters are possible. This excludes the possibility of an entirely separate second encounter of the objects as they orbit round the AGN, which is discussed in \citet{Li_and_lai_2022}.

\subsection{Physical dissipation sources}
\label{sec:mechanisms}
\subsubsection{Measurement of dissipation rates}

To uncover the physical origin of the results in Figure \ref{fig:dE_first_enc_fit}, we consider the cumulative energy dissipation per unit mass during the encounter period for the three energy dissipation terms, defined and calculated identically to our last paper \citetalias{Rowan2022} as:
\begin{enumerate}
    \item \textbf{SMBH interaction}, $\varepsilon_{\rm SMBH}$ - energy per unit time lost or gained by the binary system through shearing tidal forces induced by the SMBH.
    \item \textbf{Gas gravitational dissipation}, $\varepsilon_{\rm grav}$ - from the gravitational interaction with the surrounding gas.
    \item \textbf{Accretion}, $\varepsilon_{\rm acc}$ - due to conservation of mass and linear momentum of accreted gas particles onto the BHs. 
\end{enumerate}
These dissipation mechanisms are all calculated by taking the dot product of the velocity differential $\boldsymbol{v}_{1}-\boldsymbol{v}_{2}$ with the differential accelerations $\boldsymbol{a}_{1}-\boldsymbol{a}_{2}$ due to the force in question, as in eq.~\eqref{eq:dissipation_general_form}.
\begin{equation}
    \varepsilon = \frac{d}{dt}\bigg(\frac{E_{\rm bin}}{\mu}\bigg) = (\boldsymbol{v}_{1}-\boldsymbol{v}_{2})\cdot(\boldsymbol{a}_{1}-\boldsymbol{a}_{2}) - \frac{G\Dot{M}_{\rm bin}}{\|\boldsymbol{r}_{1}-\boldsymbol{r}_{2}\|}.
    \label{eq:dissipation_general_form}
\end{equation} 
Where the second term on the right-hand side is the change in specific energy per unit time due to any mass change in the binary, $\Dot{M}_{\rm bin}$ from accretion.

From this general expression, the dissipation due to the SMBH potential is given by
\begin{equation} 
\varepsilon_{\rm SMBH} =(\boldsymbol{v}_{1}-\boldsymbol{v}_{2})\cdot
(\boldsymbol{a}_{1,\rm SMBH}-\boldsymbol{a}_{2,\rm SMBH}),
\label{eq:work_SMBH}
\end{equation}
where the acceleration of BH $i=(1,2)$ induced by the BH is
\begin{equation}
\boldsymbol{a}_{i,\rm SMBH} = -GM_{\rm SMBH}\frac{(\boldsymbol{r}_{i}-\boldsymbol{r}_{\rm SMBH})}{||\boldsymbol{r}_{i}-\boldsymbol{r}_{\rm SMBH}||^{3}}\,.
\end{equation}
The dissipation by the local gas gravity follows the same form, now summing over gas particles (or more specifically the SPH kernels used in the tree algorithm)
\begin{equation} 
\varepsilon_{\rm grav} =(\boldsymbol{v}_{1}-\boldsymbol{v}_{2})\cdot
(\boldsymbol{a}_{1,\rm gas}-\boldsymbol{a}_{2,\rm gas}),
\label{eq:work_gas}
\end{equation}
where the acceleration of BH $i=(1,2)$ due to the gas, from all $N_{\rm p}$ SPH particles of mass $m_{\rm p}$ and position $\boldsymbol{r}_{p}$  is
\begin{equation}\label{eq:work_grav}
\boldsymbol{a}_{i,\rm gas} = -\sum_{p=1}^{N_p} Gm_{p}\frac{(\boldsymbol{r}_{i}-\boldsymbol{r}_{p})}{||\boldsymbol{r}_{i}-\boldsymbol{r}_{p}||^{3}}\, .
\end{equation}
 The accretion dissipation $\varepsilon_{\rm acc}$ is the work done on the binary due to the linear momentum of the accreted gas. Upon accretion of a particle by a BH, the BH's mass, position, velocity and acceleration are evolved by a mass weighted average of any $N_{\rm acc}$ accreted particles over the timestep
\begin{align}
    \Delta \boldsymbol{a}_{i} &= \frac{
    M_{i}\boldsymbol{a}_{i} + m_{\rm p}\sum_{j}^{N_{\rm acc}}\boldsymbol{a}_{\rm p,j}}{M_{i}+N_{\rm acc}m_{\rm p}}-\boldsymbol{a}_{i}\,.
    \label{eq:accacc}\\
    \Delta\boldsymbol{ v}_{i} &= \frac{M_{i}\boldsymbol{v}_{i}+m_{\rm p}\sum_{j}^{N_{\rm acc}}\boldsymbol{v}_{\rm p,j}}{M_{i}+N_{\rm acc}m_{\rm p}}-\boldsymbol{v}_{i}\,.
    \label{eq:accvel}
\end{align}
Where $\boldsymbol{a}_{i}$, $\boldsymbol{v}_{i}$ are the accelerations and velocities of the $i=(1,2)$ satellite BHs before the accretion event during the current timestep. $\Delta \boldsymbol{a}_{i}$ and $ \Delta{\mathbf  v}_{i}$ are the changes in their acceleration and velocity due to the impulsive momentum transfer upon accretion, where $a_{i}$ in equations \eqref{eq:accacc} and \eqref{eq:accvel} are the BH accelerations and velocities just prior the accretion impulse calculation (i.e after $a_{i,\rm SMBH}$ and $a_{i,\rm gas}$ have been incorporated into $a_{i}$). The quantities $\boldsymbol{a}_{\rm p,j}$ and $\boldsymbol{v}_{\rm p,j}$ are acceleration and velocity of accreted SPH particles. Hence, from eqns.~\eqref{eq:accacc} and \eqref{eq:accvel} the dissipation due to accretion is
\begin{equation}
    \varepsilon_{\rm acc} \approx(\boldsymbol{v}_{1}-\boldsymbol{v}_{2})\cdot(\boldsymbol{a}_{1,\rm acc} - \boldsymbol{a}_{2,\rm acc}) - \frac{G\dot{M}_{\rm bin}}{\|\boldsymbol{r}_1-\boldsymbol{r}_2\|}.
%label{eq:work_acc}
\end{equation}
where
\begin{equation}
    \boldsymbol{a}_{i,\rm acc}=\bigg(\Delta{\boldsymbol{a}_{i}}+\frac{\Delta{\boldsymbol{v}}_{i}}{\Delta t}\bigg),
\label{eq:work_acc}
\end{equation}
With the total energy dissipation as
\begin{equation}
    \centering
    \varepsilon = \varepsilon_{\rm SMBH}+\varepsilon_{\rm grav}+\varepsilon_{\rm acc}\,.
    \label{eq:diss_tot}
\end{equation}
If one integrates $\varepsilon$ over some time window, this gives the cumulative or `net total' energy exchange from the binary over the specified time window. We note that the integration matches the net change in the binary energy exactly as we deconstruct the dissipation sources and record their instantaneous value at \textit{every} timestep in the simulation, so no inaccuracies arise due incomplete sampling of the timesteps.

\subsubsection{Distinguishing orbit families}
\label{sec:families}

We differentiate between three different types of encounters by separating them into three families based on the form of their trajectories, determined by their impact parameters $p$, shown in Figure \ref{fig:enc_families}. In the co-rotating frame of the inner satellite, the first family consists of first encounters with low impact parameters where the approaching outer BH is perturbed to the left of the inner BH into a prograde orientation, i.e. orbiting in the same direction as around the SMBH. We will refer to this family as \textit{leftsided} encounters (LS for short).\footnote{the satellites execute a left-handed turn with respect to their common centre of mass (COM), $[(\boldsymbol{r}_{\rm out}-\boldsymbol{r}_{\rm in})\times(\boldsymbol{v}_{\rm out}-\boldsymbol{v}_{\rm in})]\cdot[(\boldsymbol{r}_{\rm SMBH}-\boldsymbol{r}_{\rm in})\times(\boldsymbol{v}_{\rm SMBH}-\boldsymbol{v}_{\rm in})]$<0} Further, encounters with intermediate impact parameters where the approaching outer BH initially passes on the inside (right) of the inner BH and is dragged into a retrograde orbit, which we label \textit{rightsided} (RS) encounters.\footnote{the satellites execute a right-handed turn in this case} The third family takes an initially similar approach as the RS trajectories except they turn back on themselves and reverse to a prograde orbit, i.e their trajectories are deflected by at least $180^{\circ}$ in the final lead up to the encounter. These we label as \textit{turnaround} (TA) encounters. 

These three encounter types were also identified in \citet{Boekholt_2022}. However, since there was no dissipation mechanism in that study, the binaries were loosely bound with low angular momentum and could periodically switch encounter family. The complexity of the three-body problem and this ability to switch encounter families gave rise to a fractal structure in the number of encounters as a function of impact parameter. In \citet{Boekholt_2022}, three distinctly separated islands in $p$ allow for temporary binary formation, which we do not find here, instead we observe one large formation window as a function of impact parameter. Though binaries may flip orientations afterwards, the first encounter trajectories of the binaries in \citet{Boekholt_2022} obey the same three family classifications and rough position in the space of $p$. However the size of the successful capture region of each family is significantly smaller. More specifically, the $RS$ encounters only result in binaries near the second trough at higher $p$ in the curve of $r_{\min,1}$ vs $p$, where we find encounters lead to binaries in the whole range between the two minima. Additionally the LS and TA encounters of the gasless study have their regions of possible temporary binary formation at higher $r_{\min,1}$ and are much narrower in $p$. Hence, the inclusion of gas smooths out the troughs and blends the three possible windows for formation into one larger one. The three encounter families are separated by two direct collision trajectories at the two troughs in $r_{\min,1}$ vs $p$. We can conclude that the complexity is removed as there is little to no changing of orbital families (and hence prograde/retrograde orientation) even after just two orbits as the gas is efficient at altering the energy and angular momentum of the binary, shown already in our previous paper \citetalias{Rowan2022}.

\subsubsection{Time evolution of the dissipation rates}

\begin{figure}
    \centering
    \includegraphics[width=8cm]{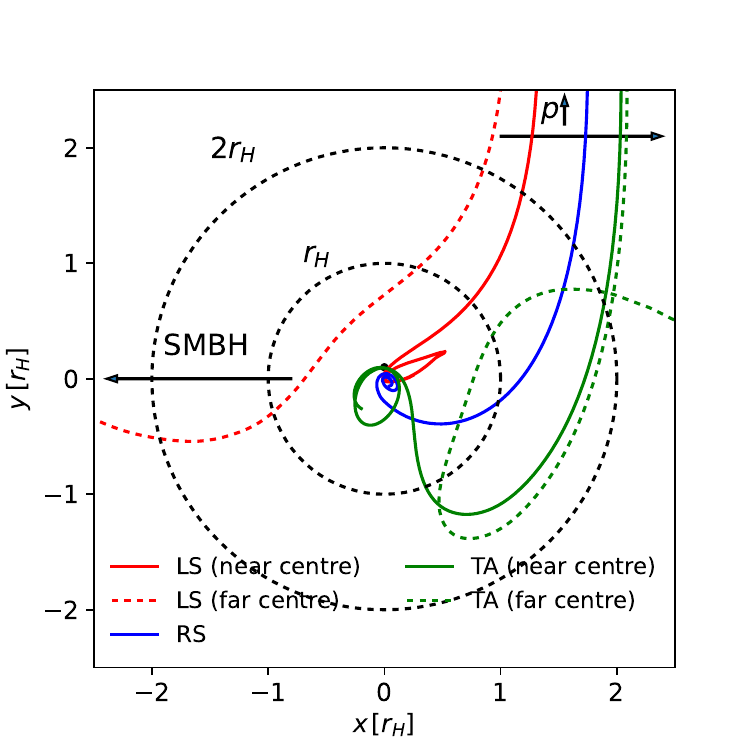}
    \caption{Classification of the trajectories of the first encounters between two sattelites moving on initially Keplerian circular orbits around a central SMBH. The trajectory of the outer satellite is shown in the frame of the inner satellite. A leftsided encounter is shown in red, rightsided in blue and turnaround in green. The dashed lines indicated the two failed encounters adjacent to each side of the capture window, to visualise how encounters far from the centre (far centre) of the capture window proceed. Their specific position on the impact parameter space is highlighted in Figure \ref{fig:cum_energy_dissipation_fid}.}
    \label{fig:enc_families}
\end{figure}

\begin{figure*}
    \centering
    \includegraphics[width=13cm]{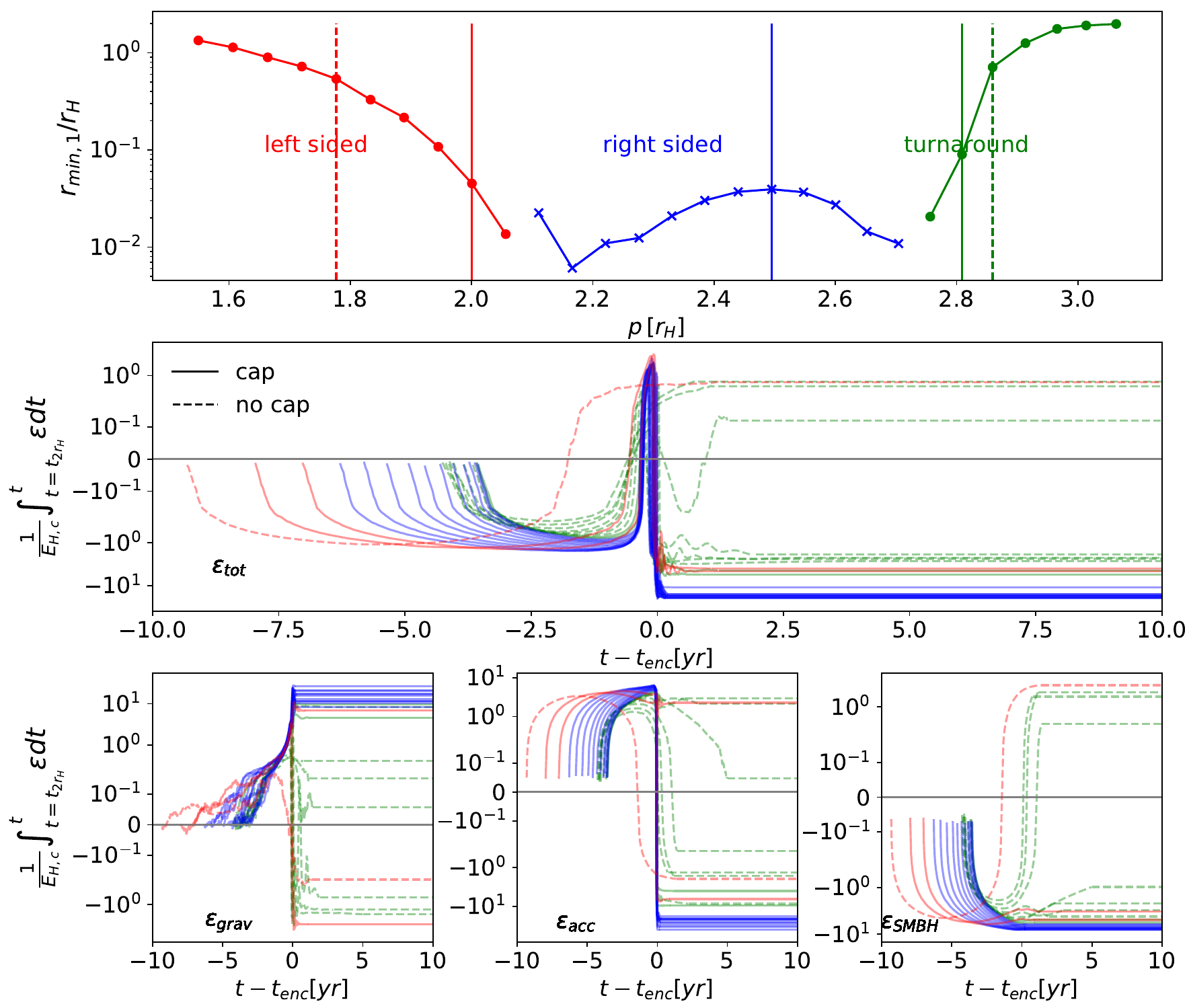}
    \caption{Cumulative energy change in all our binaries from the fiducial run during the first encounter from $2\Hillradius$ to apoapsis/exiting of Hill sphere. Results are colour-coded by encounter family; leftsided (red), rightsided (blue) and turnaround (green) encounters. The top panel re-illustrates which encounters belong to which family in the simulation suite. The dashed vertical lines indicate the two failed encounters adjacent to each side of the capture window and the bold lines highlight the three other selected models from each family used in Figure \ref{fig:enc_families} using the same colour coding. The middle panel shows the total cumulative energy change for all fiducial models. The bottom row shows the cumulative energy transfer from the local gas gravity, accretion and SMBH respectively from left to right. We artificially extend the final cumulative value from the end point of the first encounter right up to the ten year mark to more easily compare the net cumulative energy dissipation across models. Additionally, we denote unsuccessful captures with dashed lines and successful captures with solid lines. From the colour-coded results, rightsided encounters lead to the most reliably efficient binary formation which we attribute to the encounter depth dependent dissipation of Figure \ref{fig:dE_first_enc_fit}. .}
    \label{fig:cum_energy_dissipation_fid}
\end{figure*}

Figure \ref{fig:cum_energy_dissipation_fid} shows the orbit families in the parameter space of $p$ using the same colour code as in Figure \ref{fig:enc_families} and compares their cumulative energy dissipation during the first encounter.
As indicated by the top panel of Figure \ref{fig:cum_energy_dissipation_fid}, the two large troughs in the encounter depth naturally mark the transition between each family and also between prograde and retrograde binary formation, where RS encounters are retrograde and LS and TA encounters are prograde. The first trough (on the left) locates the boundary in the space of $p$ for LS and RS encounters and the second trough marks the boundary between RS and TA encounters. At these troughs is also where theoretically one can expect infinitely close encounters or direct \textit{collisions} if the impact parameters are fine tuned. Since there is a switch between prograde and retrograde orientations, there is naturally a possible set of orbits at minima of the troughs with zero angular momentum approach with an eccentricity of unity. These extremely close encounters are far better resolved in the gasless 3-body simulations. As shown in \citet{Boekholt_2022} which considers the problem in a three-body framework, it is possible to fine tune the initial conditions to have arbitrary close encounters at these two impact parameters and at others for subsequent encounters provided numerical sampling is not a limitation. 

The middle row of Figure \ref{fig:cum_energy_dissipation_fid} shows that the family of the encounter can be a good indicator of the type of energy exchange the binary will experience. RS encounters typically dissipate the most orbital energy when compared to leftsided and turnaround encounters. In all cases there is an initially positive contribution from accretion and gravitational torques as the BHs approach separations of $2r_{\mathrm{H}}$, as observed in \citetalias{Rowan2022}, which is found to be a result of a gas pileup ahead of each BH on their approach. This is initially overpowered by the shearing force on the binary from the SMBH until the final year before the closest approach, where $\varepsilon_{\rm grav}$ dominates, driving the spike in $\varepsilon_{\rm tot}$ just prior to the first closest approach ($t-t_{\rm enc}=0$) before $\varepsilon_{\rm acc}$ with some addition from $\varepsilon_{\rm SMBH}$ quickly overpower the gas gravitational forces, removing energy from the binary rapidly. Note that ultimately $\varepsilon_{\rm acc}$ and $\varepsilon_{\rm SMBH}$ have often the same order of magnitude, see further discussion in Sec.~\ref{sec:Accretionless encounters}.

Deconstructing the dissipation contributions per orbital family, we find that RS encounters typically dissipate the most energy and hence are most favourable for binary formation due to both the strong accretion and the work done by the SMBH. Interestingly, the time-evolution of the RS encounters are very consistent across all dissipation mechanisms, having net positive energy transfer from the gas to the satellites, an initially positive but rapidly negative contribution from accretion, and a negative contribution from the SMBH. Both the LS and TA encounters have a range of contributions, both positive and negative, from each dissipation mechanism. Simulations with impact parameters close to the RS encounter window exhibit the same positive energy dissipation as the RS encounters. LS and TA encounters near the deep troughs in the top panel Figure \ref{fig:cum_energy_dissipation_fid} can therefore dissipate energy as efficiently as RS encounters, but only in this small region in the space of $p$. As they deviate to higher or lower impact parameters from the central RS encounter region, dissipation gradually becomes less significant and flips when the impact parameter $p$ is sufficiently high ($\gtrsim2.8$) or low ($\lesssim1.6$) for our parameters. This is explained by the coupling of $\Delta E_{\rm{ bin}}$ with the close approach depth $r_{\rm{min},1}$ and the behaviour of $r_{\rm{min},1}$ with $p$. Where the binaries with the closest approaches, occurring in the RS encounter region, dissipate the most energy. Binaries outside the RS encounter regions have increasingly distant close approaches and so dissipate less energy.

In the wings of the encounter outside the RS region, gas gravity becomes more negative (removing energy from the binary) while dissipation from accretion becomes more positive due to less accretion during the closest approach and dissipation from the SMBH becomes positive. The reversal of the SMBH dissipation predominantly applies to the turnaround encounters as after they initially pass each other, SMBH shear flips from slowing down the binaries' approach, to accelerating them after the first periapsis passage. This is because the shear of the SMBH acts along the direction of the radial vector from the SMBH to the binary COM. As the binary approaches this shear acts against the binary motion, removing the relative kinetic energy of the binary. After a turnaround binary executes its first periapsis, the binary separation is increasing along this same vector, so the shear leads to an increase in their velocities and thus their relative energy, i.e doing positive work.

\subsubsection{Torques during the encounter}
\begin{figure*}
    \centering
    \includegraphics[width=13cm]{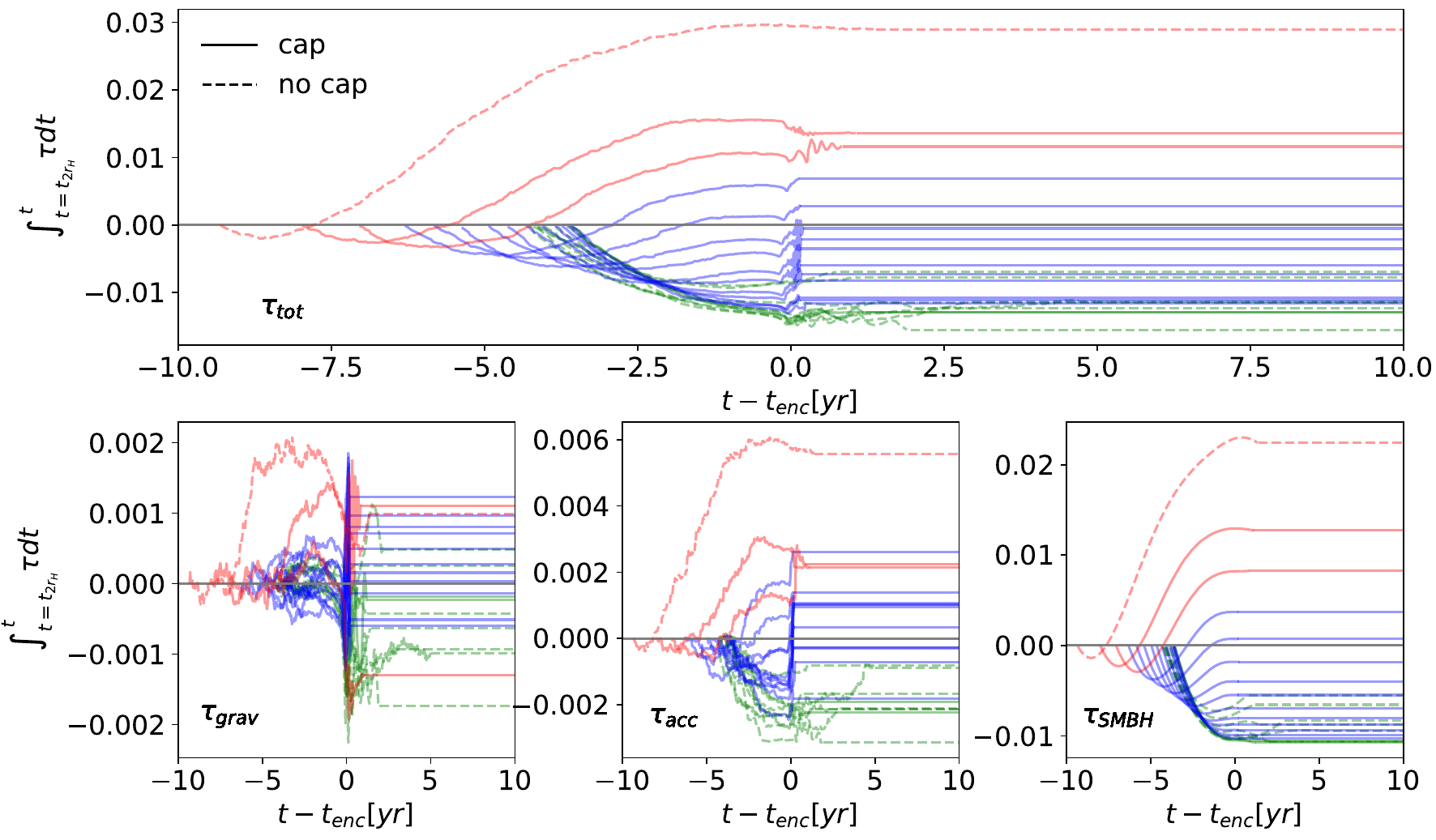}
    \caption{Cumulative torque/angular momentum change in all our binaries during the first encounter from $2\Hillradius$ to apoapsis/exiting of Hill sphere. Results are colour-coded by encounter family; leftsided (red), rightsided (blue) and turnaround (green) encounters. The top panel shows the total cumulative angular momentum change for all fiducial models. The bottom row shows the cumulative energy transfer from the local gas gravity, accretion and SMBH respectively from left to right. We artificially extend the final cumulative value from the end point of the first encounter right up to the ten year mark to more easily compare the net cumulative torque across models. Additionally, we denote unsuccessful captures with dashed lines and successful captures with solid lines. From the colour-coded results, the torque switches sign smoothly when transitioning from leftsided to turnaround encounters. This results from the SMBH torque's dependence on the angle between the vectors from the SMBH and the binary COM and from one satellite to the other.}
    \label{fig:cum_torque_fid}
\end{figure*}

Similarly to the dissipation, the strength and sign of torques induced on the binary depend on the impact parameter and family of the encounter. The cumulative torques over the first encounter (Figure \ref{fig:cum_torque_fid}) show that the net torque moves from positive to negative values as $p$ increases.
In terms of encounter families, LS encounters experience net positive torques, TA experience net negative torques and the net torque of RS encounters transitions smoothly from positive to negative. Due to the already highly eccentric nature of the binaries, we find torques highly insignificant for predicting the binary outcomes as the binary motion is almost entirely radial and so the torques $(\boldsymbol{r}_1-\boldsymbol{r}_2)\times(\boldsymbol{v}_1-\boldsymbol{v}_2)$ are minimal for the first encounter.
\subsection{Accretionless encounters}\label{sec:Accretionless encounters}
\begin{figure}
    \centering
    \includegraphics[width=9cm]{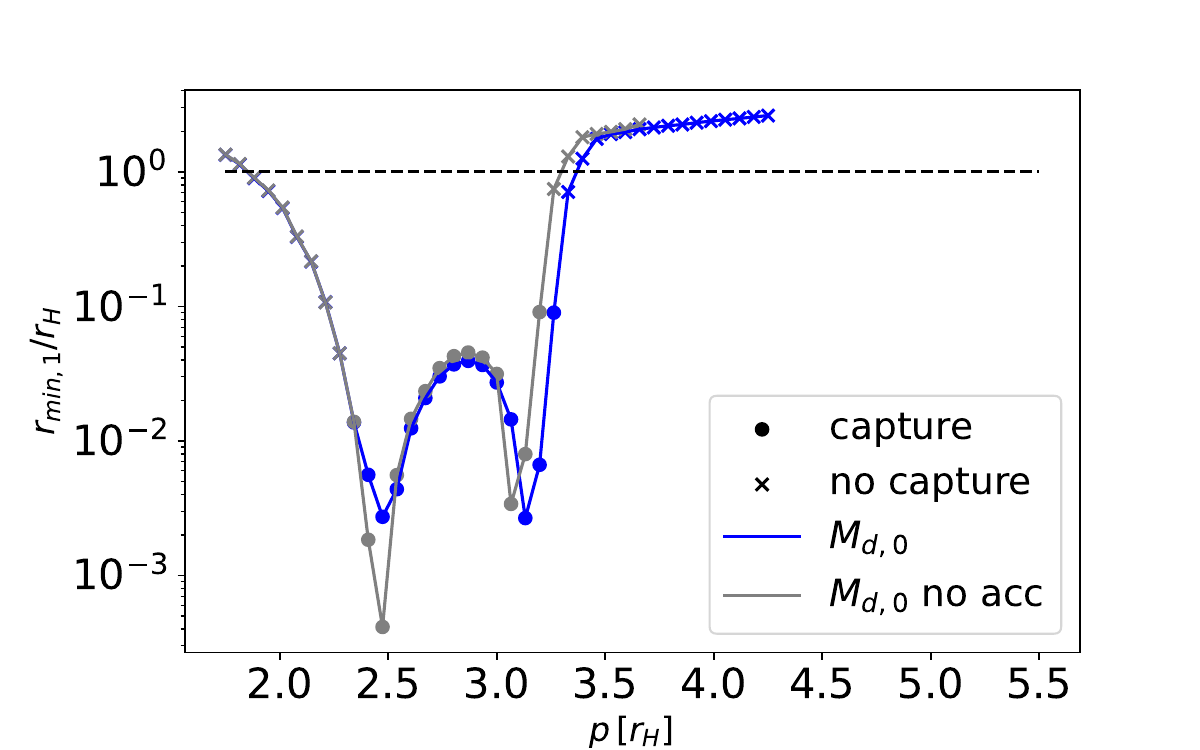}
    \caption{Minimum separation for the first approach for our fiducial run (blue) and the same run where accretion is turned off when the BH satellites get within 2 Hill radii of each other (grey) as a function of the initial radial separation $p$. Horizontal line indicates the Hill radius size of one of the objects. The outcome of all the simulations remains unchanged after switching off accretion.}
    \label{fig:noacc}
\end{figure}

In this work and in \citetalias{Rowan2022} accretion appears to dissipate energy on the same order of magnitude as gravitational gas drag. Here we test whether captures may still occur in the absence of accretion during the close encounter. Figure \ref{fig:noacc} shows the separation vs time for our fiducial run along side 30 resimulated runs of each simulation where accretion is switched off when the BHs have a separation less than $2\Hillradius$. 
The softening radius of $0.01\Hillradius$ is left unchanged. In switching off accretion at this point we preserve the initial encounter trajectory and immediate pre-encounter mass of the fiducial models so they can be directly compared. 

Figure \ref{fig:noacc} indicates that removing accretion leads to qualitatively identical outcomes with all successful binaries in the fiducial simulation suite also successfully forming in the accretionless suite and vice versa for unsuccessful captures. In addition to the scattering outcome, the depth of the initial encounter is largely unchanged, aside from the location of the two large troughs where accretionless encounters have deeper encounters for the assumed particular values of $p$. This is likely only coincident with the sensitivity of the depth with $p$ at these two turnaround points between prograde and retrograde encounters, where it was shown in \citet{Boekholt_2022} that fine tuning $p$ can lead to extremely deep encounters for only small changes in $p$. Hence the small variations in the trajectory due to the exclusion of accretion can lead to very different depths at these two key points in the parameter space of $p$. The fact the encounter depths change minimally is unsurprising as although accretion is shown to dominate overall, this is only due to a sudden reversal at the first periapsis passage, $t-t_{\rm enc}\approx 0$. Thus the encounter trajectory is only perturbed by the SMBH and gas gravity prior to the close encounter, so the encounter depth remains largely unchanged. The capture cross section for accretionless encounters is then calculated to be $\lambda_{\rm noacc}=0.92\Hillradius$ in the range $2.34r_{\mathrm{H}}<p<3.26r_{\mathrm{H}}$, identical to the fiducial model.

\begin{figure*}
    \centering
    \includegraphics[width=13cm]{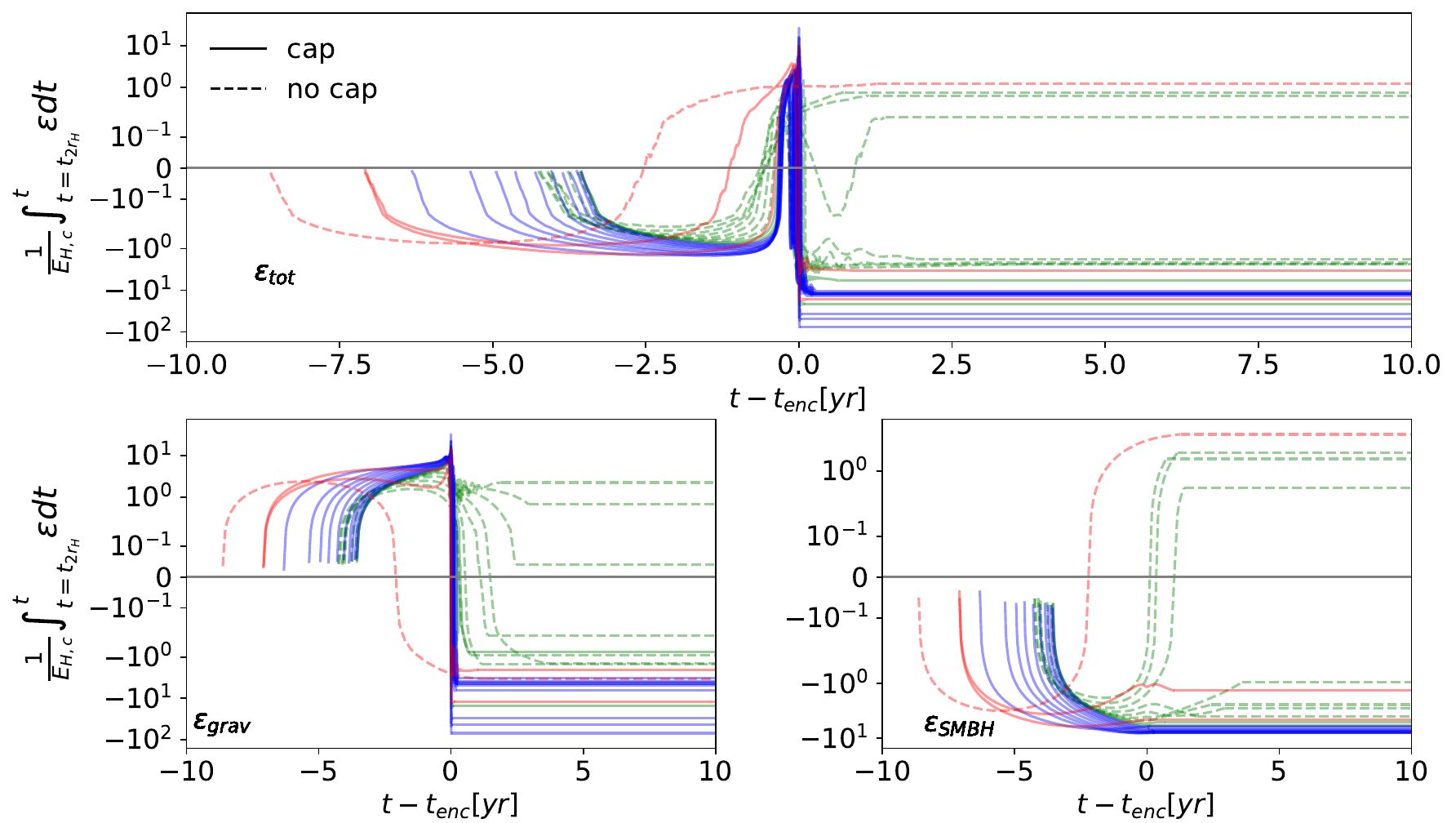}
    \caption{Cumulative energy change change in all our binaries from the no-accretion run during the first encounter from $2\Hillradius$ to apoapsis/exiting of Hill sphere. Results are colour-coded by encounter family; leftsided (red), rightsided (blue) and turnaround (green) encounters. The top panel shows the total cumulative energy change for all fiducial models. The cumulative energy transfer from the local gas gravity and SMBH are shown respectively on the left and right of the second row. We artificially extend the final cumulative value from the end point of the first encounter right up to the ten year mark to more easily compare the net cumulative dissipation across models. Additionally, we denote unsuccessful captures with dashed lines and successful captures with solid lines. Turning off accretion leads to the replication of its effects in the dissipation from the local gas gravity.}
    \label{fig:diss_and_torque_noacc}
\end{figure*}

\begin{figure}
    \centering
    \includegraphics[width=9cm]{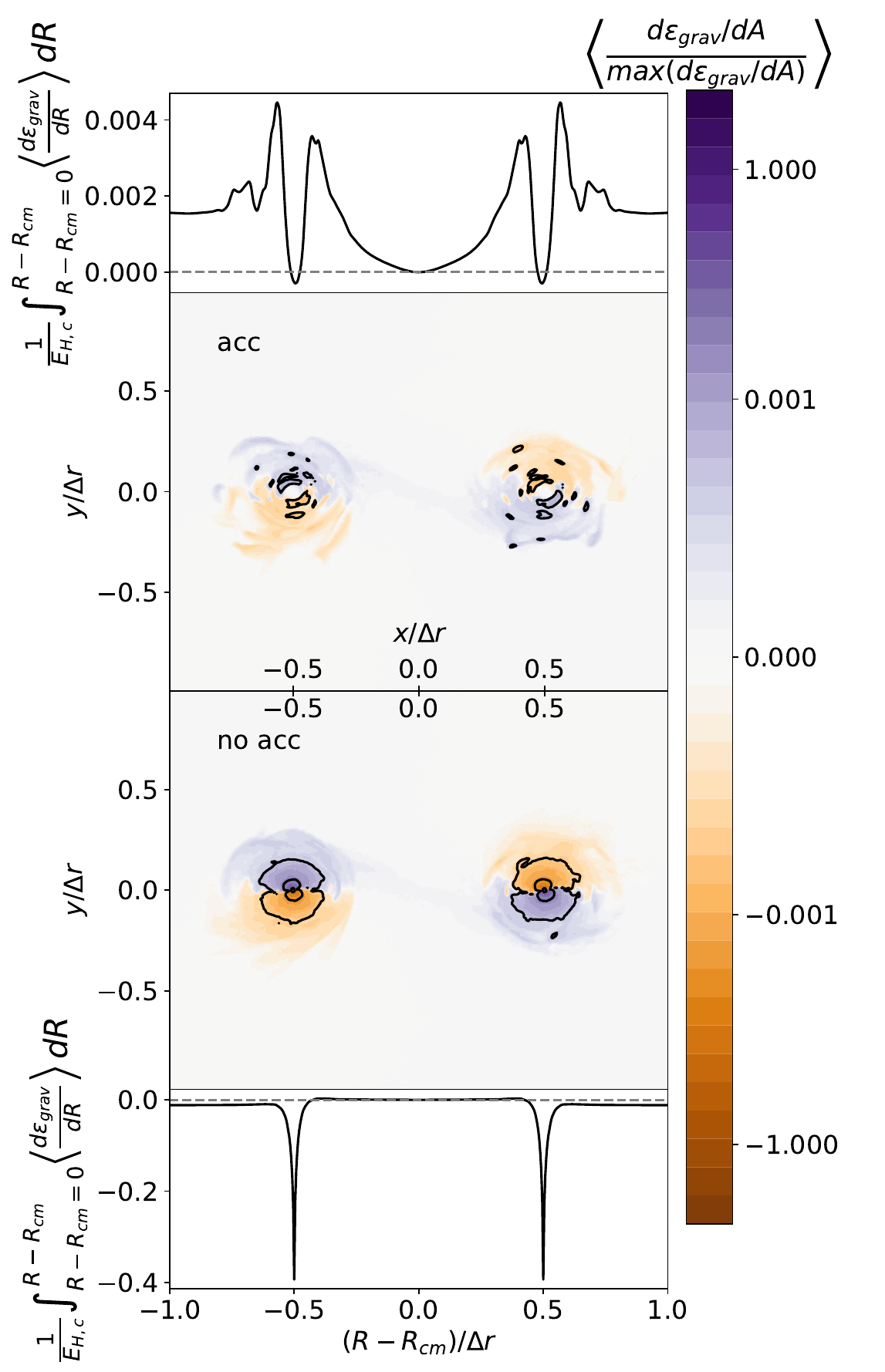}
    \caption{\textit{Central two panels}: 2D average dissipation per area from gravitational interaction with the local gas \eqref{eq:work_grav} over the first encounter for a representative accretion-enabled (top) and accretionless (bottom) simulation with identical impact parameters. The map is centred on the centre of mass $R_{\rm{CM}}$ of the binary and distances normalised to the current separation of the binary $\Delta r$. The dissipation is normalised to the max of both simulations to compare the relative strengths. The black lines show equal contours in each simulation, showing the enhanced dissipation close to the BHs in the accretionless case. \textit{Top panel}: Average gravitational energy dissipation integrated from the $x-y$ centre of mass radially outwards in the plane of the binary for the accretion-enabled case. \textit{Bottom panel}: Same as the top panel but for the accretionless model. The radial and 2D maps are aligned so they can be directly compared along their axes. The binaries execute their orbit in a counter clockwise direction.}
    \label{fig:dissmaps}
\end{figure}

The non-negligible effect of removing accretion is in the behaviour of $\varepsilon_{\rm grav}$ immediately after $t-t_{\rm enc}\approx0$, shown in Figure \ref{fig:diss_and_torque_noacc}. This occurs during the time frame we would expect the extremely negative energy dissipation from accretion at the first periapsis passage, $t-t_{\rm enc}\approx 0$.
When accretion is removed, the time evolution of $\varepsilon_{\rm acc}$ in the fiducial simulations is reproduced in $\varepsilon_{\rm grav}$, demonstrating a highly efficient period of energy removal from the binary after $t-t_{\rm enc}\approx0$, driving the cumulative dissipation to negative values. More formally, the time evolution of $\varepsilon_{\rm grav}$ without accretion mimics the rapid transition from positive to negative values observed in $\varepsilon_{\rm acc}$ in our accretion simulations. This is more akin to the findings of \citet{Li_Dempsey_Lai+2022} where gravity acts to add energy prior to periapsis then reverses afterwards, resulting in a net energy removal. This is not present when accretion is enabled because accretion can alter the gas morphology close to the binary. When accretion is switched off, gas that would normally be accreted upon entering $r_{\rm acc}$ can instead pass around the BH. Alternatively, in the frame of the BH, the accretion headwind felt by the BH can now pass directly through the accretion boundary and accumulate behind the BH. To visualise this, we calculate a surface density plot of the dissipation rate, averaged over the first encounter, from the local gas gravity $<d\varepsilon_{\rm grav}/dA>$ and show this alongside the radially integrated cumulative dissipation over the first encounter period in Figure \ref{fig:dissmaps} for an accretionless and accretion enabled simulation case.

Figure \ref{fig:dissmaps} shows, as expected, that the morphologies are largely similar prior to the closest approach in the simulations with and without accretion. In the accretionless simulation there is a density buildup within the softening radius of the gas's gravitational effect on the satellites, $r_{\rm soft}$ though this is largely axisymmetric and does not strongly affect the binary trajectory at this moment. Just before and after the periapsis passage there are dense
%enhanced 
gas clumps behind each BH in the accretionless encounters where the gas has passed around the BH and accumulated behind them, similar to the wakes caused by dynamical friction. 

\begin{figure}
    \centering
    \includegraphics[width=8cm]{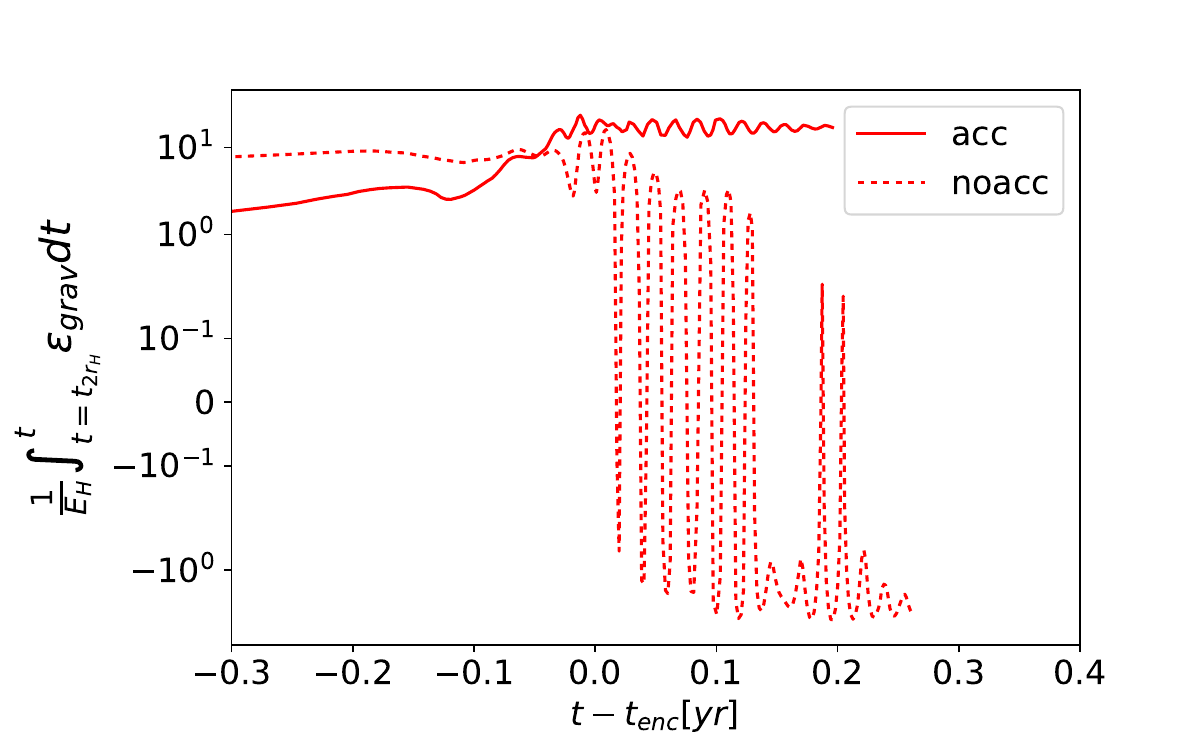}
    \caption{A zoom-in on the cumulative energy change of the simulations with gas accretion being enabled and turned off, respectively, in Figure \ref{fig:dissmaps} around the first periapsis passage. Artificially turning accretion off leads to far stronger oscillations in the work done on the binary, due to the stronger gravitational tug of the gas minidiscs.}
    \label{fig:dissipation_squiggles}
\end{figure}

Figure \ref{fig:dissipation_squiggles} shows a zoom-in on the cumulative work done by the gas gravity on the satellites in the simulations with and without gas accretion near the first periapsis passage. The dissipation from $\varepsilon_{\rm grav}$ oscillates violently over the close encounter period. 
This was first observed in \citetalias{Rowan2022} and is a result of the formation of non-axisymetric, highly dense, inhomogeneities in the CSMDs from the BH tidal forces. As they orbit their respective BHs, the clumps tug on the BHs, resulting in a periodic oscillation in $\varepsilon_{\rm grav}$. We find the oscillations have an orbital frequency which is akin to a Keplerian orbit of $\sim 0.4 r_{\min,1}$. Though the strength of the oscillations in the binary energy warrants a discussion of the possibility they may have their own gravitational wave imprint on the encounter in a similar manner to other gas-induced GW signitures suggested in the literature \cite[e.g.][]{Kocsis+2011,Yunes+2011,Hayasaki2013,Barausse+2014,Barausse+2015,Cole2023,Nouri2023}. However the characteristic frequency of the oscillations is typically of the order $\sim10^{-7}-10^{-6}$Hz which is unfortunately too low to be measured even with LISA.

We conclude that the formation of these inhomogeneities is due to the tidal warping of the CSMDs being maximal at the periapsis of the encounter, this is further evidenced by the fact they form exactly during this part of the binary orbit (see Figure \ref{fig:dissipation_squiggles}). The characteristic orbital radius of the inhomogeneities is therefore unsurprising as the BHs do not pass within $0.5r_{\min,1}$ of each others CSMDs, so the gas in these regions is better retained around their original BH following periapsis. Thus gas can remain at high densities within distances smaller than the periapsis separation (i.e $|R-R_{1}| < 0.5r_{\min,1}$ for the inner BH). 

The oscillating dissipation phenomenon from the inner CSMDs is present in both simulations with and without accretion, however the oscillations are larger in the latter case. The cumulative effect of the rapidly varying $\varepsilon_{\rm grav}$, is a net dissipation in the binary energy by the time of the first apoapsis passage. In the simulations with accretion, this is net positive in most cases, i.e. the binary gains energy and becomes less bound, in agreement with our previous results in \citetalias{Rowan2022}. In contrast, in our accretionless simulations, the net dissipation is significantly negative. This is the result of the allowed permeation of what would have been the accretion radius in the previous accretion enabled simulations by the gas.  
Gas that would normally accrete \textit{against} the BHs motion (see Figure \ref{fig:cum_energy_dissipation_fid}) and produce the strong negative dissipation instead accumulates behind the BH. While the CSMDs perturbations are still present, this permeation leads to a net removal of binary energy via gravitational drag. 

\begin{figure*}
    \centering
    \includegraphics[width=13cm]{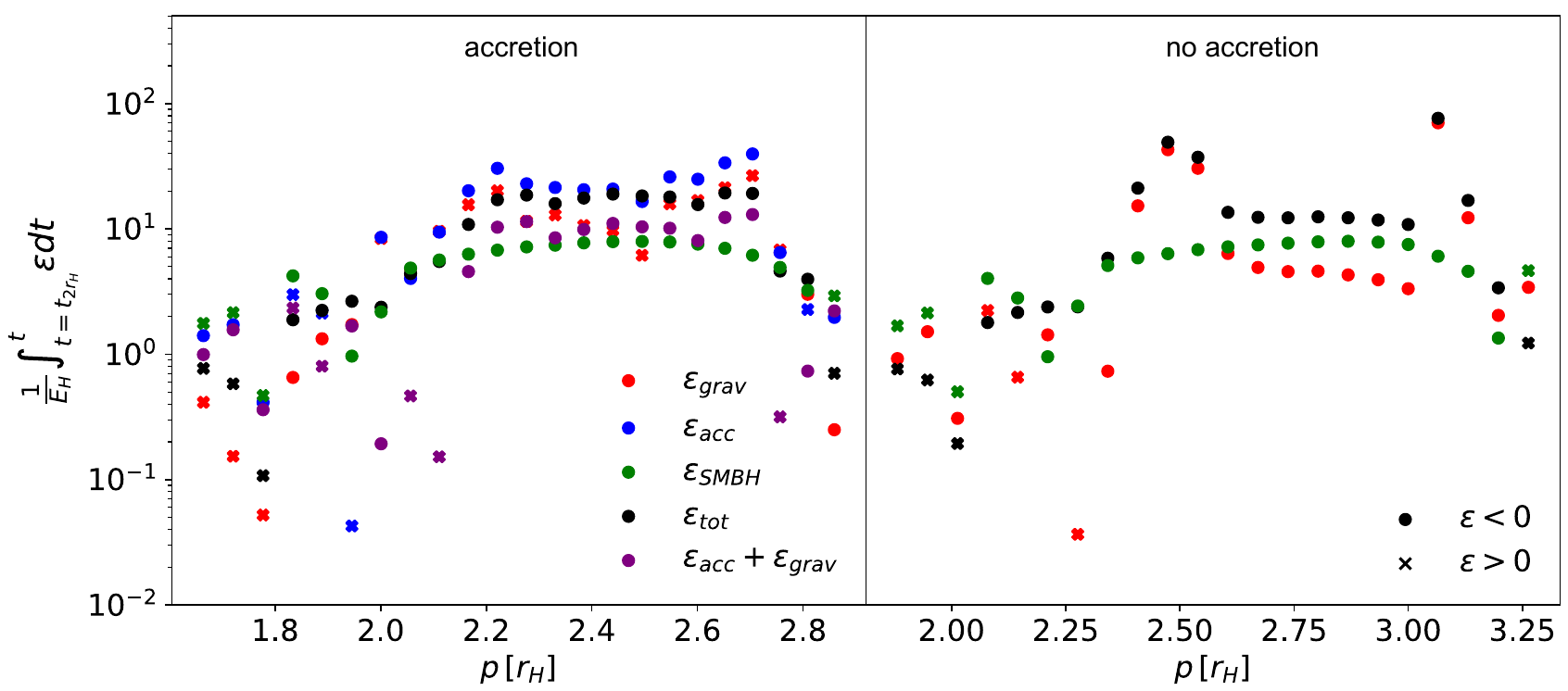}
    \caption{Net energy change of the binaries over their first encounters, comparing the fiducial simulations where gas accretion is possible (left) to the simulations where accretion is turned off (right). The net contributions from different physical processes are shown with different colours.  In the accretion-enabled panel, we also show the strength of $\varepsilon_{\rm grav}+\varepsilon_{\rm acc}$ to compare with $\varepsilon_{\rm grav}$ in the accretionless simulations. When accretion is turned off, the sign of the net gravitational dissipation switches from positive to negative. Though accretion is the most efficient remover of energy, its omission only changes the dissipation by roughly a factor two. This indicates that it fairly accurately models the net dissipation from $\varepsilon_{\rm acc}$ contributions near the accreting boundary. In otherwords one may neglect accretion and retain fairly similar results qualitatively.}
    \label{fig:dissipation_by_mechanism_noacc}
\end{figure*}

Figure \ref{fig:dissipation_by_mechanism_noacc} directly compares the net contribution of various physical processes to the binary energy dissipation during the first encounter. 
We find that artificially removing the accretion leads to qualitatively similar behaviour, with the RS encounters dissipating the most energy, where now it is dominated by $\varepsilon_{\rm grav}$ and $\varepsilon_{\rm SMBH}$ instead of $\varepsilon_{\rm grav}$ and $\varepsilon_{\rm acc}$. While enabling accretion significantly alters the impact parameter dependence of the integrated $\varepsilon_{\rm grav}$, we find that the relative strength of the gas effects is similar across the range of impact parameters when comparing $\varepsilon_{\rm acc}+\varepsilon_{\rm grav}$ to $\varepsilon_{\rm grav}$ in simulations with and without accretion. This suggests that although the dominant dissipation term shifts from gas gravity to accretion when accretion is enabled, it preserves the overall expected time-dependent variance and strength of the overall dissipation at least to within a factor of few.

\subsection{Dependence on different accretion disc densities}
\label{sec:diff_density}
\begin{figure}
    \centering
    \includegraphics[width=9cm]{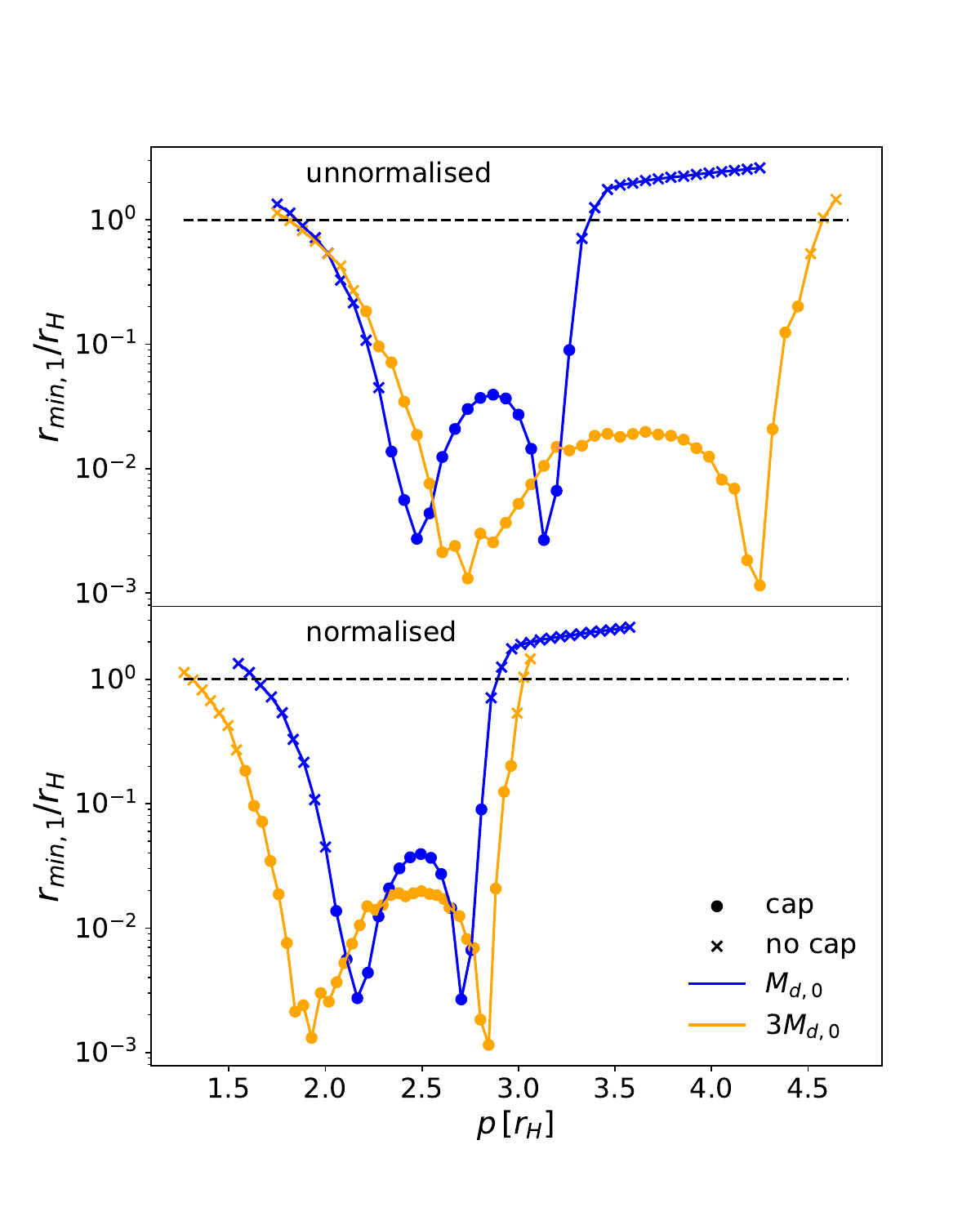}
    \caption{Minimum separation for our fiducial run (blue) compared to the $3M_{\mathrm{d},0}$ run (orange) for the first approach as a function of the initial radial separation $p$. On the top row is the unaltered initial form of the curve. On the bottom we show the same two curves, where the minimum separations are now normalised to the size of the Hill sphere when the binaries reach $2r_{\mathrm{H}}$ in separation. The normalisation process is described by eq. \eqref{eq:normed_cross_sect}.}
    \label{fig:high_dens_window}
\end{figure}

We now consider BH scatterings in a 3 times higher density environment than in our fiducial model but otherwise identical initial conditions so that the disk mass is $M_{\mathrm{d}} = 3M_{\mathrm{d},0}$. This includes the modelling of accretion as in the fiducial simulations. This suite is comprised of 45 simulations, 6 more than the fiducial run to span the larger space of impact parameters for hard encounters.
Figure \ref{fig:high_dens_window} compares the encounter depth vs impact parameters. The window of captures in $p$ is greatly enhanced with additional gas.
This matches the conclusion of \cite{DeLaurentiis2022}, \cite{Li_Dempsey_Lai+2022} and our sibling paper; \citet{Henry_inprep}. The added gas also has the effect of shifting the 'valley' of the encounter depths to higher impact parameters, this explains why for our high disc densities in \citetalias{Rowan2022} we found very different encounters compared to the lower mass simulations. There are two effects causing this shift, one is the increased mass buildup in the discs, which means the BHs are perturbed earlier along their orbit since the disc mass can effectively be added to the BH mass when they are sufficiently far away. The second is that accretion causes the Hill sphere of the objects to increase on their approach to each other. We remove the accretion dependence of the window in the bottom panel by normalising the cross sections, which are presented in units of the Hill sphere, to the Hill sphere of the BHs when they intersect twice each others Hill radii, $r_{\rm{H_{2H}}}$, instead of their initial Hill radii, $r_{\rm{H_{t=0}}}$, and denote these cross sections with a tilde:
\begin{equation}
    \Delta\tilde{\lambda}_i = \frac{1}{2}\left(p_{\rm{i+1}}\frac{r_{\rm{H_{t=0}}}}{r_{\rm{H_{2H}}}} - p_{\rm{i}}\frac{r_{\rm{H_{t=0}}}}{r_{\rm{H_{2H}}}}\right)
    \label{eq:normed_cross_secti}
\end{equation}
\begin{equation}
    \tilde{\lambda}\left( N_e \geq 2 \right) = \sum_i \frac{1}{2}\Delta\tilde{\lambda}_i I_{[N_e \geq 2]}(N_{e,i}),
    \label{eq:normed_cross_sect}
\end{equation}
This centres the windows perfectly, showing that accretion is responsible for the shifting of the window. However, the overall window of the encounters is still larger, where the two troughs are further apart as well as the RS and turnaround encounter regions occur at lower and higher impact parameters, respectively. Additionally, captures successfully take place at shallower initial encounter depths, even when $r_{\min,1} > 0.1$ unlike in the fiducial model. This evidence points to encounters in AGN discs with higher densities being more favourable for binary formation; allowing binaries more sparsely separated to have stronger encounters and affording binaries with weaker encounters a greater chance for formation. 

\begin{figure*}
    \centering
    \includegraphics[width=13cm]{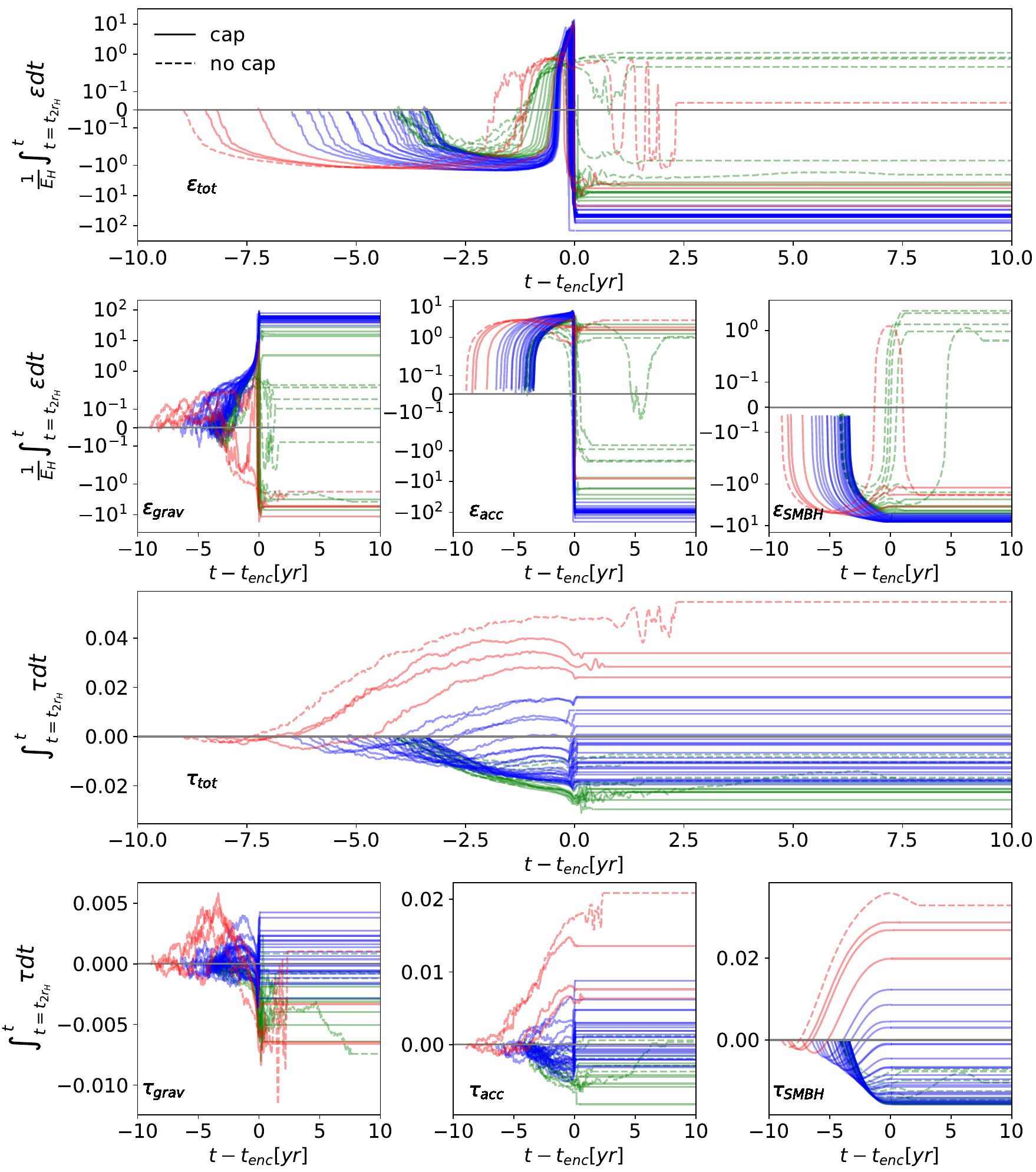}
    \caption{Cumulative energy and angular momentum change in all our binaries from the run with three times the AGN disc density ($3M_{\rm d,0}$) during the first encounter from $2\Hillradius$ to apoapsis/exiting of Hill sphere. Results are colour-coded by encounter family; rightsided (red), leftsided (blue) and turnaround (green) encounters. The top panel shows the total cumulative energy change for all fiducial models. The cumulative energy transfer from the local gas gravity, accretion and SMBH are shown respectively from left to right of the second row. The cumulative torque is shown on the 3rd row, with the breakdown from the three dissipation mechanisms similarly displayed in row 4. We artificially extend the final cumulative value from the end point of the first encounter right up to the ten year mark to more easily compare the net cumulative torque across models. Additionally, we denote unsuccessful captures with dashed lines and successful captures with solid lines. The enhanced gas density does not change the form of the dissipation or torques but increases their strengths.}
    \label{fig:diss_and_torque_3Md}
\end{figure*}

\begin{figure*}
    \centering
    \includegraphics[width=13cm]{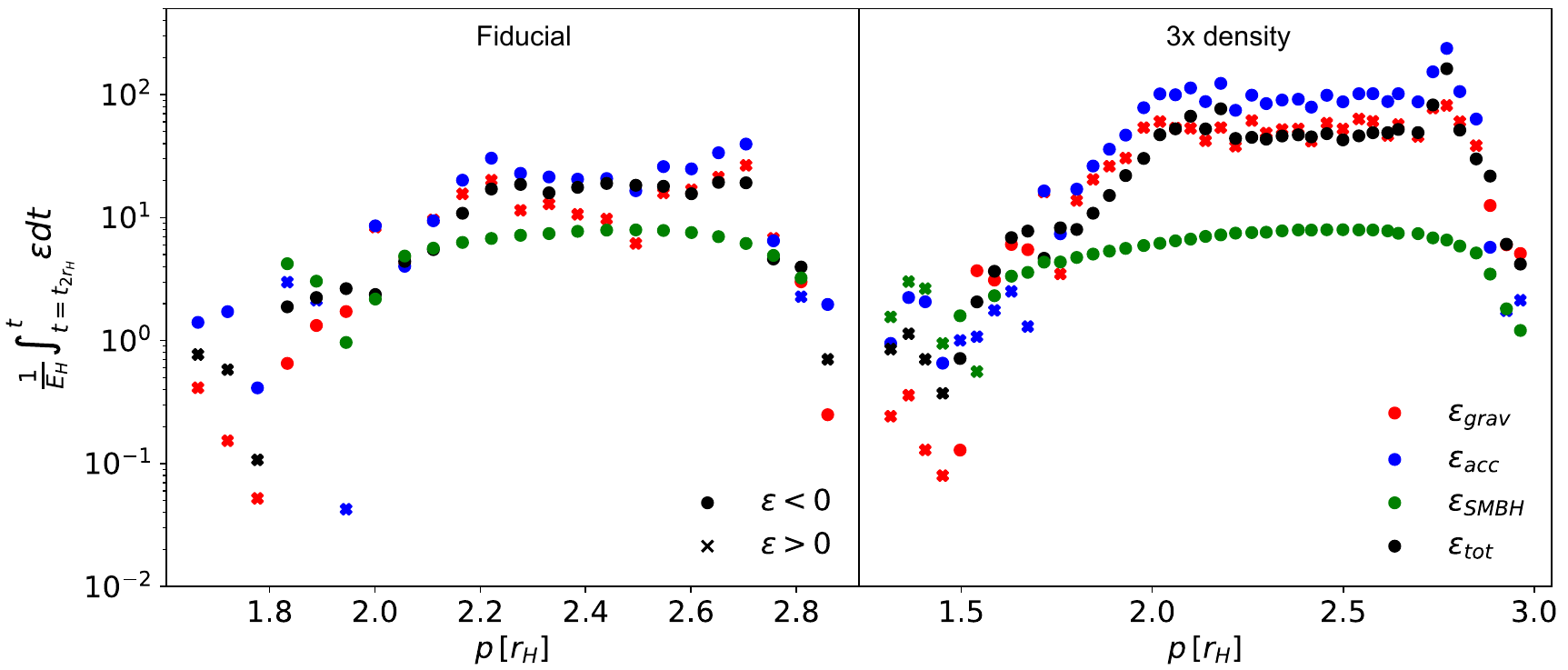}
    \caption{Net energy change of the satellites over their first encounters, comparing the fiducial simulations (left) to the $3M_{\rm d,0}$ enhanced disc density simulations (right). Also shown are the net contributions from each dissipation mechanism. The enhancement of the dissipation from the increased ambient density is clear in the rightsided encounters (see Figure~\ref{fig:cum_energy_dissipation_fid} for definition) in the central flat plateau of the encounter window, showing on average approximately three times greater energy dissipation.}
    \label{fig:dissipation_by_mechanism_3Md}
\end{figure*}

Figure \ref{fig:diss_and_torque_3Md} shows the contributions of the different physical processes to the satellite energy dissipation and the torques over the first encounter and per source over time.
Qualitatively, the overall trend is the same initially. Dissipation is initially negative due to the SMBH, before rapidly peaking positive just prior to encounter due to gas gravity followed by a sharp drop due to strong negative dissipation from accretion. The torque behaviour is also preserved from the fiducial model with the SMBH dominating the torque in the negative direction for nearly all models, while the gas gravity and accretion induces negative torques in the turnaround encounters, gradually flipping to positive torque across the window of leftsided encounters and inducing significantly positive torques for RS encounters. While the behaviour of the dissipation mechanisms is unchanged from the fiducial model, the strength of the dissipation and torques from both accretion and gas gravity are around a factor of $3\times$ larger (see Figure \ref{fig:dissipation_by_mechanism_3Md} for clarity).
For the torques this is less impactful as the SMBH still dominates by a factor of a few. However the dissipation becomes enhanced by this  factor of five as it is dominated by gravity and accretion, indicating that more energy can be dissipated in higher gas density environments. Note that this is independent of any enhanced mass gain by the BHs as our energy unit normalisation $E_{\rm \mathrm{H},c}$ and its reduced mass $\mu$ dependence is calculated for each model. The capture cross section reflects this favourability, with a value of $\Tilde{\lambda}_{3M_{\rm d}}=1.591$. In Table \ref{tab:lambda} we summarise the normalised and unnormalised cross sections from each of our simulation suites.
\begin{table}
    \centering
    \begin{tabular}{|c|c|c|c|} 
    \hline
    model & $\lambda$ & $\Tilde{\lambda}$ & $\tilde{\lambda}/\tilde{\lambda}_{fid}$ \\
    \hline
    fiducial & $0.98\pm0.13$  & $0.86\pm0.11$ & 1 \\
    no gas & $0.0407\pm0.0011$  & $0.0407\pm0.0011$ & $1.00\pm0.13$  \\    
    $3\times$ gas density & $2.30\pm0.19$ & $1.59\pm0.13$ & $1.81\pm0.28$\\
     no-accretion & $0.92\pm0.19$  & $0.80\pm0.13$ & $0.93\pm0.25$ \\
    \hline

    \end{tabular}
    \caption{1D cross sections $\lambda$ in the space of the impact parameter $p$ as calculated by eq. 
    \eqref{eq:cross_sect} for successful binary formation. These values quantify formally the range in $p$ that permits binary formation. Shown from left to right are the different simulation suites, the standard cross section $\lambda$, the accretion normalised cross sections $\tilde{\lambda}$ (see eq. \ref{eq:normed_cross_sect}), and the ratio of the normalised cross sections to that of the fiducial simulation suite.} 
    \label{tab:lambda}

\end{table}

\begin{figure}
    \centering
    \includegraphics[width=9cm]{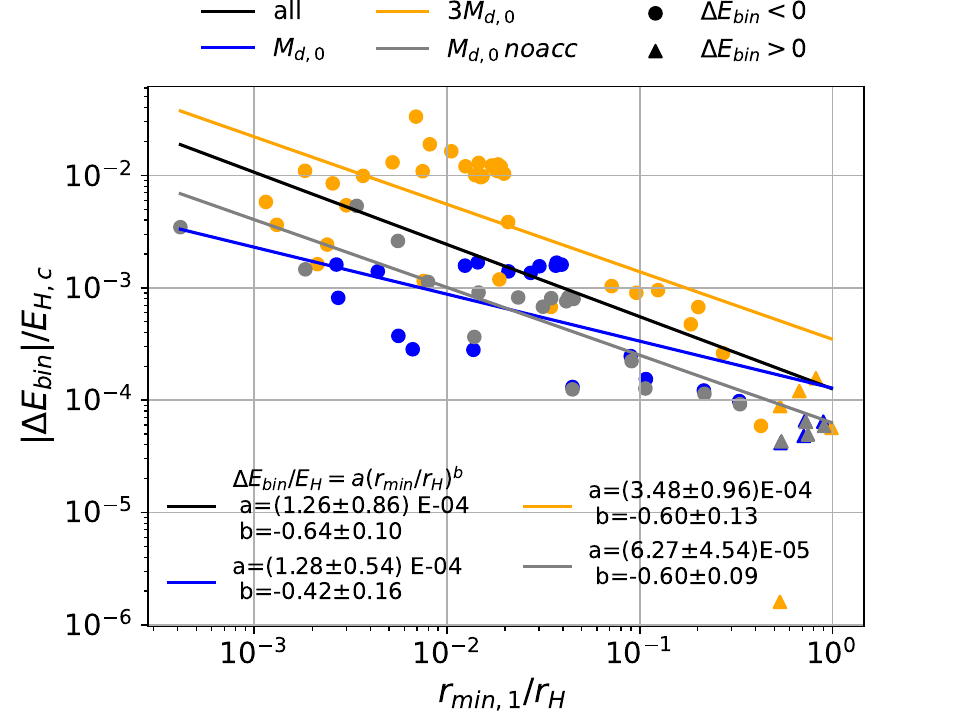}
    \caption{Energy change in the binaries, normalised to $E_{\rm \mathrm{H},c}$, during the first encounter as a function of the first periapsis $r_{\rm min,1}$ as in Figure~\ref{fig:dE_first_enc_fit} but for all simulations including those with an increased disk mass and those where accretion is turned off. Dotted points indicate where energy is \text{removed} to the binary while triangular points represent models where energy is \textit{added}. Only binaries that pass within $\Hillradius$ are shown. Also shown in the same colour as the raw data are the power-law fits for the fiducial $M_{\rm d,0}$ (blue), $3M_{\rm d,0}$, accretionless $M_{\rm d,0}$ and whole sample (black).}
    \label{fig:dE_first_enc_fit_by_suite}
\end{figure}

Recall that we have shown in Figure \ref{fig:dE_first_enc_fit} that the depth of the first encounter greatly affects the orbital energy dissipation of the binary for simulations with fiducial disc model with different impact parameters. Figure \ref{fig:dE_first_enc_fit_by_suite} shows the energy dissipation vs closest approach for all our simulations, including those with an increased disk mass and those where accretion is turned off. From the results, we find that compared to our fiducial model, the density enhancement and removala of accretion leads to a steeper dependence of the energy dissipated with the periapsis distance $r_{min,1}$. Additionally, if we compare the $a$ coefficients we can quantify the average energy dissipation increase with density. With a value of $3.48/1.28 = 2.69\pm1.33$ this corresponds to a scaling with the disc density $\Sigma$ of $\Delta E_{\rm bin}\propto\Sigma^{0.9\pm0.44}$ close to the prediction from our other paper \citet{Henry_inprep} that finds dissipation scales with density to the power of unity, albeit with a considerable error margin here. 

\subsection{Parameter space of captures}
While analytic and semi analytic studies of BH binary population synthesis in AGN (e.g \citealt{Miralda2000,Freitag2006,Hopman2006,Bartos2017,Rasskazov2019,Secunda2019,Tagawa2020,McKernan+2020}) are well suited for simulating the large BH population in AGN, they currently utilise very simplistic assumptions for the outcome of an encounter, oversimplifying the complex interactions with the gas. 
A revised, more physically motivated set of analytical tools for deducing the outcome of an encounter would provide far more realistic and reliable estimators for the binary formation rate which is essential for estimating the BH merger rate for GW astronomy. 
While we have numerically derived an expression relating dissipation and encounter depth, the encounter depth itself is affected non trivially by the gaseous effects. A more useful expression may be obtained if one ascertains a parameter space for captures using encounter parameters that are not largely effected by the complex gas morphology of the encounter itself. We consider two such parameter spaces, starting with the the binary relative energy at the start of the encounter $E_{2\mathrm{H}}$ at $2r_{\mathrm{H}}$ and the impact parameter of the binary at the Hill radius $p_{\rm 1\mathrm{H}}$ and . We define these quantities specifically in eq. \eqref{eq:E2H} and eq. \eqref{eq:p1H} respectively. Where eq. \eqref{eq:E2H} is equation eq. \eqref{eq:two_body_energy} evaluated when the binaries are at $2r_{H}$ separation.
\begin{equation}
    \centering
    E_{\rm 2\mathrm{H}}=\frac{1}{2}\mu \|\boldsymbol{v}_1-\boldsymbol{v}_2\|^2 - \frac{GM_{\rm{bin}}\mu}{2r_{\mathrm{H}}}\,.
    \label{eq:E2H}
\end{equation}
%
%\begin{equation}
%    \centering
%    p_{\rm 1H} = r_{\mathrm{H}}\sin{\bigg(\arccos{\bigg(\frac{(\boldsymbol{v}_2-%\boldsymbol{v}_1)\cdot(\boldsymbol{r}_2-\boldsymbol{r}_1)}{\|\boldsymbol{v}_2-%\boldsymbol{v}_1\|\,\|\boldsymbol{r}_2-\boldsymbol{r}_1\|}\bigg)}}\bigg)
%    \label{eq:p1H}
%\end{equation}
\begin{equation}
    \centering
    p_{\rm 1H} = r_{\mathrm{H}}\sqrt{1-\bigg(\frac{(\boldsymbol{v}_2-\boldsymbol{v}_1)\cdot(\boldsymbol{r}_2-\boldsymbol{r}_1)}{\|\boldsymbol{v}_2-\boldsymbol{v}_1\|\,\|\boldsymbol{r}_2-\boldsymbol{r}_1\|}\bigg)^{2}}
    \label{eq:p1H}
\end{equation}

Where as before, $\boldsymbol{r}_i$ and $\boldsymbol{v}_i$ are the positions of the satellite black holes $i=(1,2)$. $M_{\rm bin}=M_{1}+M_{2}$ is the mass of the outer binary and $\mu=M_{1}M_{2}/M_{\rm bin}$ is its reduced mass. To reiterate, eq. \eqref{eq:E2H} is evaluated when the binary separation first reduces to $2r_{\mathrm{H}}$ and eq. \eqref{eq:p1H} is evaluated when the separation reaches $r_{\mathrm{H}}$.

\begin{figure}
    \centering
    \includegraphics[width=8cm]{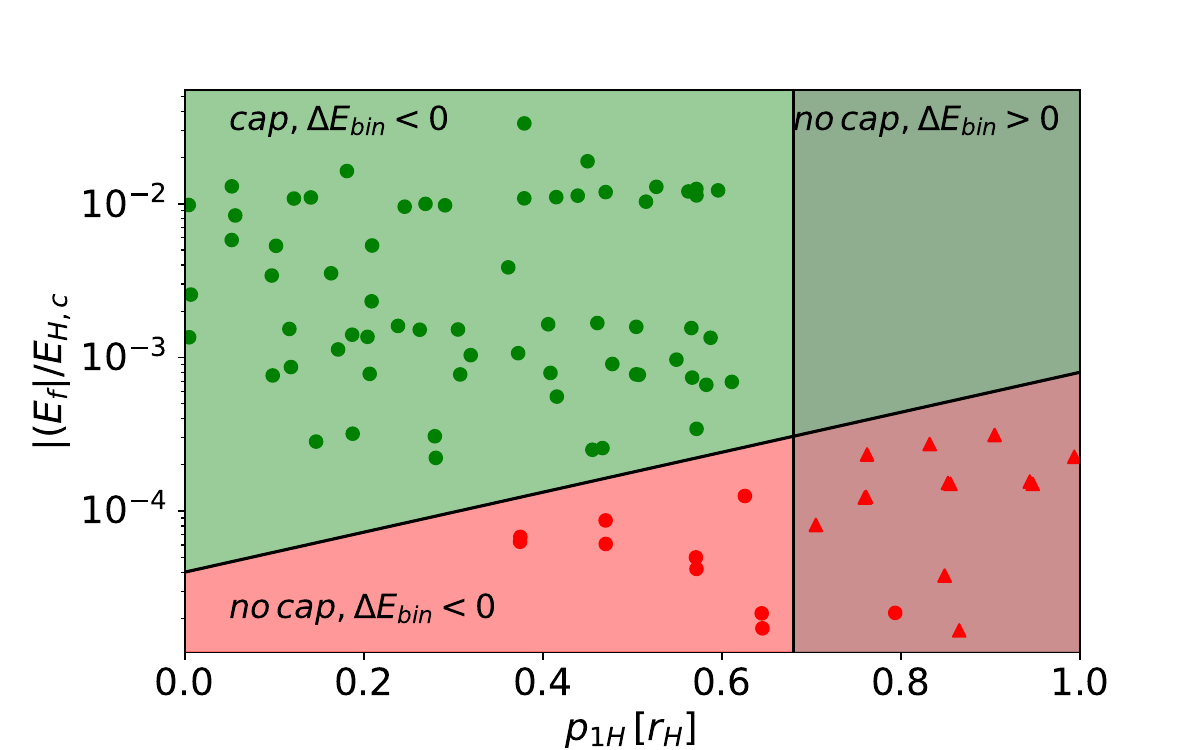}
    \caption{Orbital energy, in units of the energy of a circular orbit at the Hill sphere $E_{\rm H,c}$, of the binaries after the first encounter (either when executing first apoapsis or upon leaving the Hill sphere if unbound), $E_{\rm f}$ vs their impact parameter at one Hill radius $p_{\rm{1H}}$ as defined in \eqref{eq:p1H}. The results show three distinct regions of parameter space, 1) where binaries successfully form through sufficient energy dissipation, 2) where binaries  dissipate energy but not sufficiently to remain bound and 3) where binaries decouple through energy gained during the encounter. The function of the positively sloped line separating the successful from unsuccessful formations is the critical final energy, $E_{\rm f,crit}$, needed to remain bound, represented by the log-linear function in eq. \eqref{eq:Ef}.}
    \label{fig:param_space_p_vs_Ef}
\end{figure}

Before constructing our analytic tool for predicting the energy dissipated, we first have to understand how much energy must be lost. Figure \ref{fig:param_space_p_vs_Ef} shows the total energy of the binary after the first encounter, $E_{\rm f}$, i.e after the work done by the three dissipation mechanisms during the encounter, $\Delta E_{\rm bin}$, has been added to the initial energy at $2\Hillradius$, $E_{\rm 2\mathrm{H}}$. One might consider using the energy at $2\Hillradius$ but the impact parameter at $\Hillradius$ an odd choice for our parameterisation. However, as shown in Figures \ref{fig:cum_energy_dissipation_fid}, \ref{fig:diss_and_torque_noacc} and \ref{fig:diss_and_torque_3Md} the energy exchange of the binary becomes important at scales beyond the Hill radius. Though the dissipation from $2r_{\mathrm{H}}$ is important for determining the outcome, we find no binary formation beyond just $p_{\rm 1\mathrm{H}}\gtrsim0.68\Hillradius$. Thus in semi-analytical studies, one needs only consider binaries entering a single Hill radius and have access to $E_{\rm 2\mathrm{H}}$ and $p_{\rm 1\mathrm{H}}$ to determine the scattering outcome.

\begin{figure}
    \centering
    \includegraphics[width=8cm]{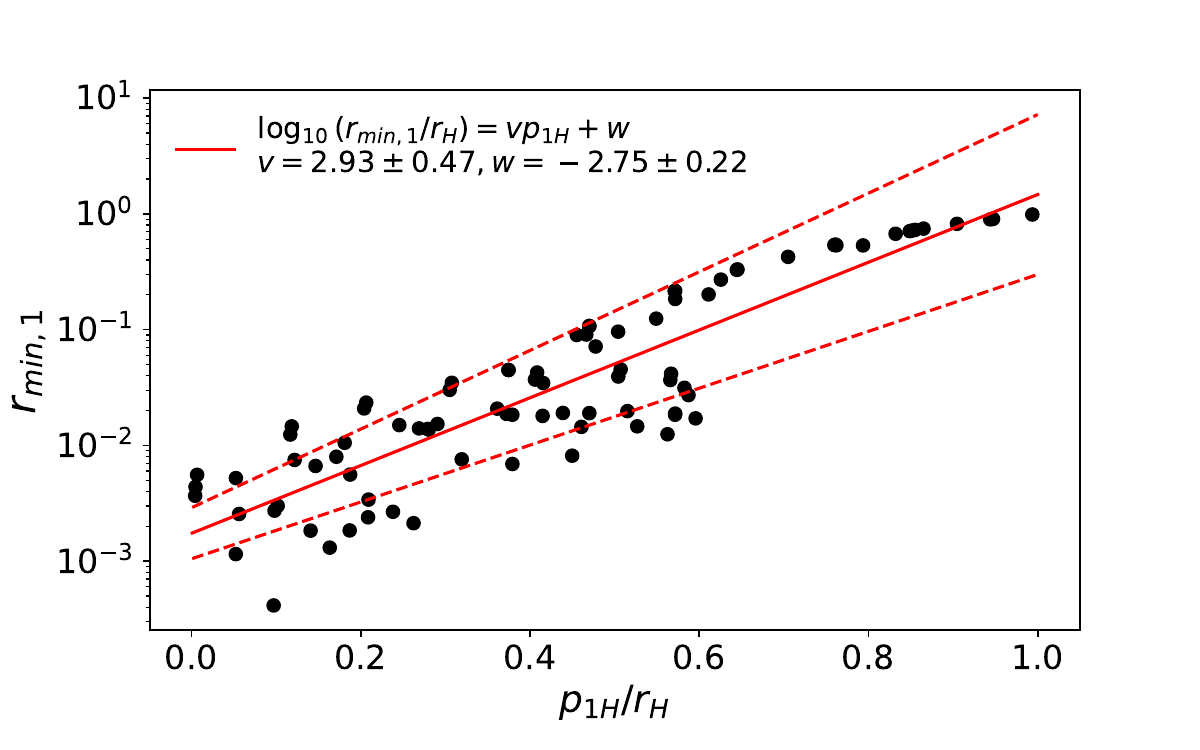}
    \caption{First encounter periapsis scattered against the impact parameter at one Hill radius $p_{\mathrm{1H}}$ (Eq.~\ref{eq:p1H}) for all simulations, showing that smaller impact parameters lead to closer encounters, with some scatter for lower values of $p_{\rm{1H}}$ where the very small periapsese becomes increasingly sensitive to small changes in energy. Also shown is the log-linear line of best fit (solid red) and its error (dashed red) according to eq. \eqref{eq:rmin_vs_p1H}.
    }
    \label{fig:impact_vs_depth}
\end{figure}

In the plot, a very clear parameter space for successful binary formation emerges. At the bottom of the parameter space there is an island of failed captures, where for $p_{\rm 1\mathrm{H}}/\Hillradius\gtrsim0.68$ energy is added to the binary. In the other failed capture region where $p_{\rm 1\mathrm{H}}/\Hillradius\lesssim0.68$, this means the binary did not dissipate enough energy to be sufficiently hard so as not to be decoupled by the SMBH. The slope in the boundary between each island implies that before first periapsis, a binary with identical initial energy must dissipate more to remain bound if its impact parameter is higher. This is no doubt due to correlation between $p_{\rm 1\mathrm{H}}$ and the encounter depth $r_{\min,1}$, which we confirm in Figure \ref{fig:impact_vs_depth}. We define this sloped boundary as the function $E_{\rm f, crit}(p_{\rm 1H})$, shown below in eq. \eqref{eq:Ef}.
\begin{equation}
    \centering
    \log_{10}(E_{\rm f,crit}/E_{\mathrm{H},c}) =  1.31p_{1\mathrm{H}}-4.34
    \label{eq:Ef}
\end{equation}
Next, we determine the scaling between $p_{\rm 1H}$ and $r_{\min,1}$, which is accurately informed by our full hydrodynamical approach and therefore includes the effects and stochasticity introduced by the gas. We show our function for $r_{\rm min,1}$ in terms of $p_{\mathrm{1H}}$ (see eq. \ref{eq:rmin_vs_p1H}) alongside the data in Figure \ref{fig:impact_vs_depth}.
\begin{equation}
    \centering
    \log_{10}(r_{\rm min,1}/r_{\mathrm{H}}) =  vp_{1\mathrm{H}}-w
    \label{eq:rmin_vs_p1H}
\end{equation}
with constants $v=2.93\pm0.47$ and $w=-2.75\pm0.22$. This relation can then be inserted into $\eqref{eq:powerlaw}$ to obtain the function $\Delta E_{\rm bin}(p_{\mathrm{1H}})$. 

Using our fits of the $\Delta E_{\rm bin} (r_{\min,1})$ relation to each simulation suite, we construct the parameter space of $p_{\rm 1\mathrm{H}}-E_{2\mathrm{H}}$ that allows for binary formation in eq. \eqref{eq:cap_criterion_simple}
\begin{equation}
    \frac{E_{2\mathrm{H}}}{E_{\rm \mathrm{H},c}} < \left(\frac{E_{2\mathrm{H}}}{E_{\rm \mathrm{H},c}}\right)_{\rm crit} = E_{\rm f,crit}(p_{\rm 1H}) - \Delta E_{\rm bin}(p_{\rm 1H})\,.
    \label{eq:cap_criterion_simple}
\end{equation}
 In simple terms, this equation calculates the critical final energy required to remain bound based on its impact parameter $p_{\rm 1\mathrm{H}}$ and subtracts how much energy it was expected to dissipate based on the same impact parameter to calculate its expected energy at $2H$. Therefore, it calculates the expected energy at $E_{\rm 2\mathrm{H}}$ that would allow a binary to remain bound for its impact parameter $p_{\rm 1\mathrm{H}}$. In explicit form, both the first and second term take on power laws. When all constants combined and simplified this gives:
\begin{equation}
    \frac{E_{2\mathrm{H}}}{E_{\rm \mathrm{H},c}} < \left(\frac{E_{2\mathrm{H}}}{E_{\rm \mathrm{H},c}}\right)_{\rm crit} = 10^{k_{1}\frac{p_{\rm{1H}}}{r_H}+c_{1}} - 10^{k_{2}\frac{p_{\rm{1H}}}{r_H}+c_{2}}\,.
    \label{eq:cap_criterion}
\end{equation}
We determine the constants for this equation for all of our simulations suites collectively and individually and display them in Table \ref{tab:fit_params}. According to our criterion, a BH-BH scattering will result in a successfully formed binary if the following is true 
\begin{equation}
  \bigg[\frac{E_{2\mathrm{H}}}{E_{\rm \mathrm{H},c}} < \left(\frac{E_{2\mathrm{H}}}{E_{\rm \mathrm{H},c}}\right)_{\rm crit}\bigg]  \wedge \bigg[p_{\rm 1H} <0.68\bigg]\,. 
\end{equation}
This new expression meets our aim of being independent of any variables that could be affected by the complex hydrodynamics of the encounter and since the energy is normalised to $E_{\rm \mathrm{H},c}$ it is also independent of the binary mass, thus is insensitive to any mass buildup due to accretion \textit{before} the encounter. We summarise the process of constructing the binary formation criterion in short steps below.
\begin{figure}
    \centering
    \includegraphics[width=9cm]{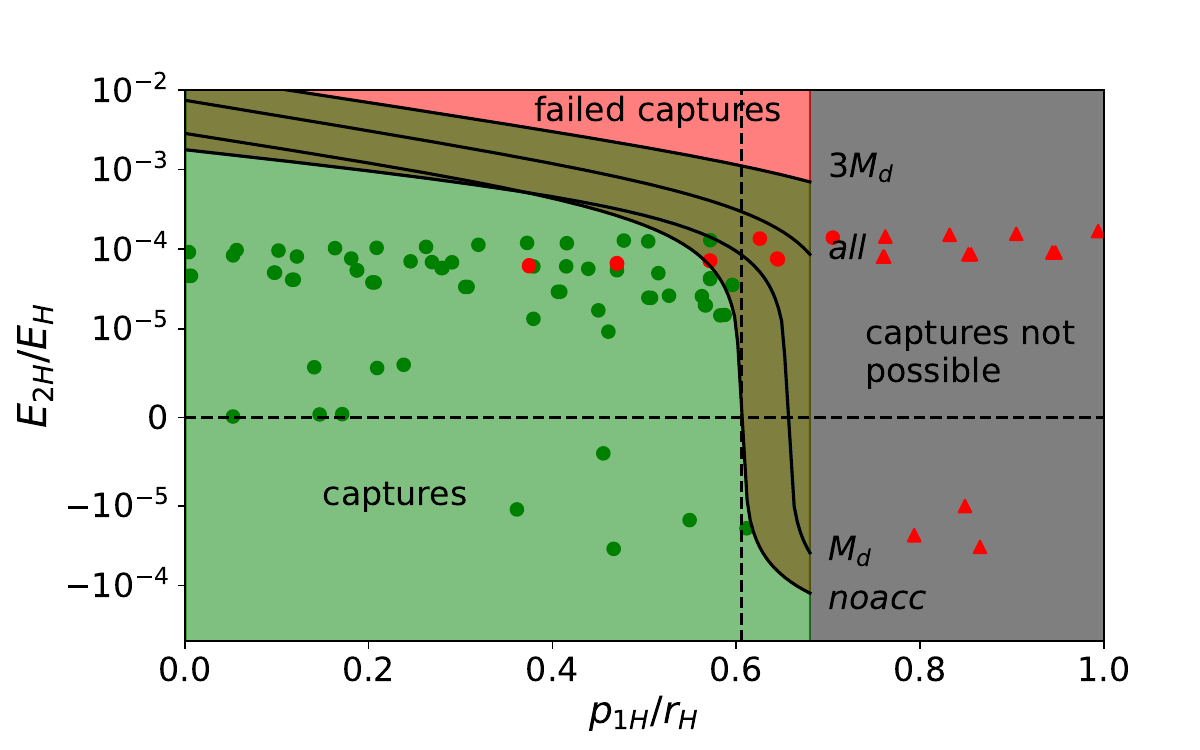}
    \caption{Encounter energy at two Hill radii, $E_{\rm 2\mathrm{H}}$, for all our models as a function of their impact parameter as measured at $p_{\rm 1H}$ overlayed on our analytically derived parameter space. The red area highlights where we expect failed captures due to a lack of sufficient energy removal. The green area is the parameter space where we expect successful captures. The grey area presents a firm barrier to capture as energy is added to the binaries in this region which don't pass deep into each others' Hill sphere. The overlapping red and green area is the area of the parameter space where the outcome depends on our simulation suite. The boundary of the successful formation region for each suite is traced by the black curves and labelled in the plot, calculated via eq. \eqref{eq:cap_criterion_simple}.}
    \label{fig:param_space_analytical}
\end{figure}
We apply our criterion to the data first in Figure \ref{fig:param_space_analytical}. The figure shows three distinct regions,
\begin{enumerate}
    \item The already known region of zero binary formation at $p_{\rm 1\mathrm{H}}\gtrsim0.68$. 
    \item The region of successful binary formation under the curve where the binary can remove enough energy to remain bound and unperturbed by the SMBH.
    \item The region above the curve where the binary was too energetic to dissipate enough energy to form a stable binary. 
\end{enumerate}
The vertical dashed line indicates the transition to where the binary must already have a negative two-body energy to remain bound for our fiducial model. The curves that separates the successful and unsuccessful captures are our criterion eq. \ref{eq:cap_criterion} calculated for each suite using the corresponding $a$ and $b$ values. We summarise the methodology of constructing our capture criterion in numbered steps.

\begin{steps}
    \item[\textbf{Summary of Binary Formation Criterion}]
    \item[]
    \item Determine the binary's two body energy at a separation of two Hill radii, $E_{\rm 2\mathrm{H}}$, using eq. \eqref{eq:E2H}.
    \item Determine the binary's impact parameter at separation of one Hill radii, $p_{\rm 1\mathrm{H}}$, using eq. \eqref{eq:p1H}.
    \item Use the impact parameter $p_{\rm 1\mathrm{H}}$ to determine the close approach distance $r_{\rm min,1}$ using eq. \eqref{eq:rmin_vs_p1H}.
    \item Determine the expected energy dissipation, $\Delta E_{\rm bin}$, with eq. \eqref{eq:powerlaw} using $r_{\rm min,1}$.
    \item Calculate the minimum energy $E_{\rm f,crit}$ required to stay bound via eq. \eqref{eq:Ef} using $p_{\mathrm{1H}}$.
    \item Determine if the energy dissipation criteria, eq. \ref{eq:cap_criterion} is satisfied using $ E_{\rm f,crit}$, $\Delta E_{\rm bin}$ and $E_{\mathrm{2H}}$. With constants for eq. \ref{eq:cap_criterion} given by Table \ref{tab:fit_params}.   
\end{steps}

While this criterion is a significant improvement, it naturally still makes assumptions. These include its assumption of only 2D encounters and equal mass ratios for the two BHs. Further work is needed to determine generalisations for non-coplanar encounters misaligned with the AGN disc, and for asymmetric mass ratios. Additionally, the relationship may be depend on the mass of the BHs, since we have not verified that the mass of the minidiscs scales one to one with the BH mass. Given a large portion of the dissipation in rightsided encounters comes from colliding minidiscs, alternate ratios of minidisc to BH masses due to differing BH masses could affect the amount of energy dissipated in the same manner as changing the ratio via changing the AGN disc density.

\begin{figure*}
    \centering
    \includegraphics[width=17cm]{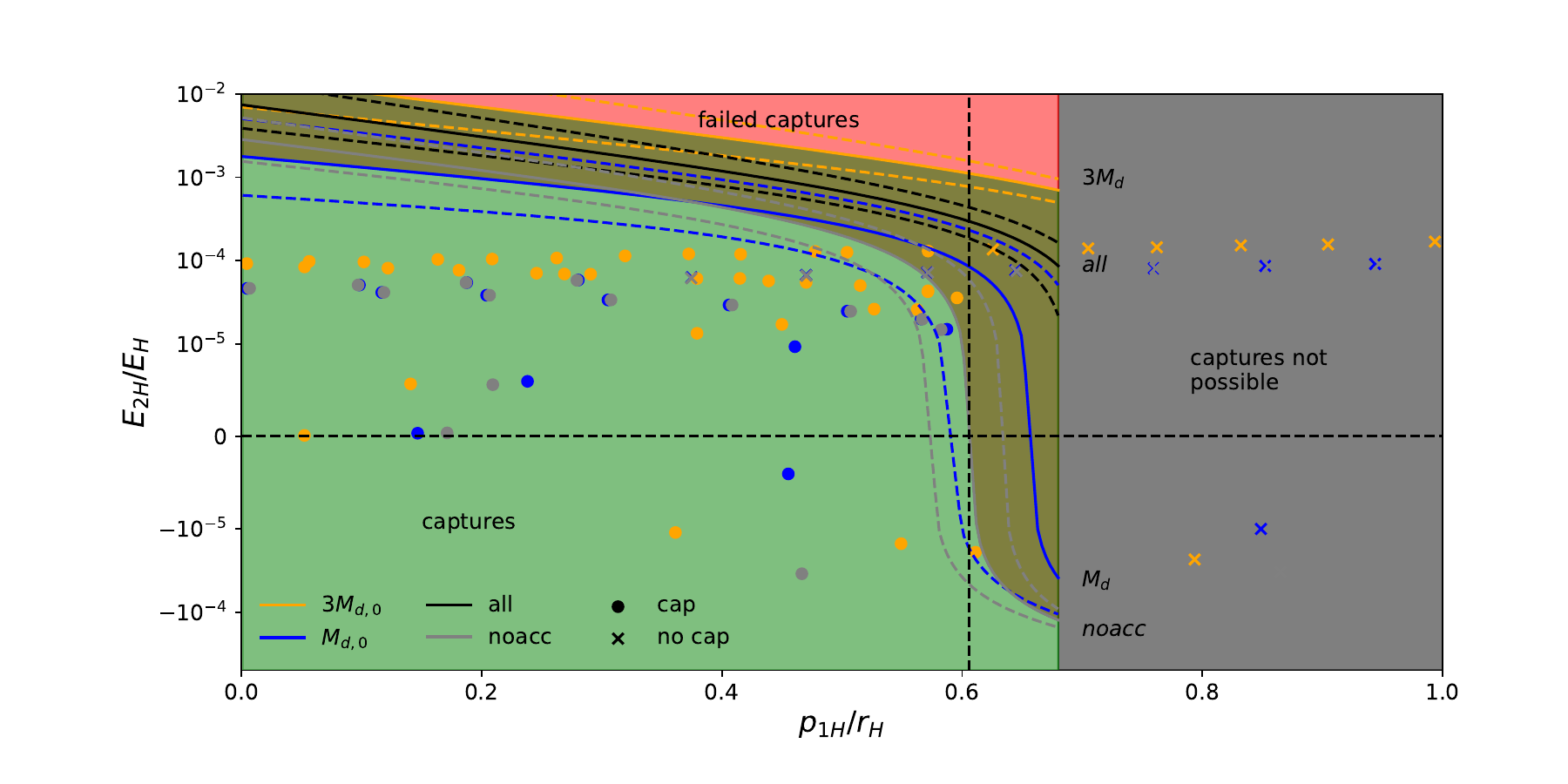}
    \caption{Encounter energy at two Hill radii, $E_{\rm 2\mathrm{H}}$, as in Figure~\ref{fig:param_space_analytical} but for all simulations including those with an increased disk mass and those where accretion is turned off as a function of their impact parameter as measured at $p_{\rm 1H}$ overlayed on our analytically derived parameter space. The red area highlights where we expect failed captures due to a lack of sufficient energy removal. The green area is the parameter space where we expect successful captures. The grey area presents a firm barrier to capture as energy is added to the binaries in this region which don't pass deep into each others' Hill sphere. The overlapping red and green area is the area of the parameter space where the outcome depends on our simulation suite. The boundary of the successful formation region for each suite is traced by the solid colour-coded curves according to the simulation suite. The dashed curves span the error range of the boundary region. Out of 115 simulations, we find only 5 outliers. We highlight the point where accretionless encounters that already have a negative two body energy still need to dissipate energy further with the horizontal and vertical dashed lines.}
    \label{fig:param_space_analytical_all}
\end{figure*}

We verify the accuracy of the criterion in Figure \ref{fig:param_space_analytical_all} by breaking it down by suite and also showing the errors in the curves. Counting the false positives and false negatives in each suite we find a predictive success rating of 94\%, which for the errors in $b$ and $a$ of eq \eqref{eq:powerlaw} is considerably good.  Additionally, though we have shown its form to be mass dependent based on the curves in Figures \ref{fig:param_space_analytical} \& \ref{fig:param_space_analytical_all}, our relation only holds for the expected mean AGN density for our parameters and our inflated 3x density model. A more formal density scaling would allow the relation to hold for arbitrary or at least a range of AGN densities, which is the subject of our sibling paper, \citet{Henry_inprep}.
\begin{table}
\begin{tabular}{|c|c|c|c|c|} 

 \hline
            & $k_{1}$ & $c_{1}$ & $k_{2}$ & $c_{2}$ \\
 \hline
 $M_{\rm d,0}$  & 1.31  & -4.34 & -1.74 & -2.52 \\ 
 \hline
 $3M_{\rm d,0}$ & 1.31  & -4.34 & -2.48 & -1.57\\
 \hline
 noacc      & 1.31  & -4.34 & -1.60 & -2.65 \\
 \hline
 all        & 1.31  & -4.34 & -2.21 & -2.02 \\
 \hline
\end{tabular}

\caption{Fit parameters for eq. \ref{eq:cap_criterion} for our $M_{\rm d}=M_{\rm d,0}$, $M_{\rm d} = 3M_{\rm d,0}$ and accretionless simulations, also shown is the average fit using all our simulations.}
\label{tab:fit_params}
\end{table}
\section{Discussion}
\label{sec:discussion}
\subsection{Implications for mergers}
Given the high efficiency with which binaries form in the simulations and that most of them are retrograde, if we combine this with the result from \citetalias{Rowan2022} that retrograde binaries have a non-negligible merger rate, it would appear that overall the AGN channel is highly conducive to black hole mergers. In reality however, there is also non-zero radial and vertical velocity dispersion in the stellar mass BH population in AGN that is not encorporated in this work. The additional vertical component will naturally lead to interactions outside of the midplane where the gaseous background is less dense, leading to a lower probability of capture as indicated here, in \citetalias{Rowan2022} and in \citet{Li_Dempsey_Lai+2022}. Additionally, little is known of what gas morphology we should expect around a satellite in an AGN disc. We expect encounters with BHs that have entered from outside or at the extremities of the AGN disc will involve less gas since they cannot replenish gas lost from their Hill sphere to the SMBH as quickly or at all prior to the encounter \citep{Kratter+2010,Kocsis+2011}. Though this additional aspect to the problem will likely reduce the chance for binary formation and mergers, it is expected that the inclination of objects will decrease significantly over the lifetime of the AGN due to dynamical friction \citep[e.g.][]{Tagawa2020,Generozov2023_disccap} between the satellites and the SMBH disc.

Based on the density dependence of the capture criterion (eq. \eqref{eq:cap_criterion}) favouring higher densities this would suggest a bias towards formations and merger of BH binaries closer to the SMBH according to eq. \eqref{eq:rho_R}. However there is the competing factor of the velocity shear of the satellites, which is more pronounced closer to the SMBH. An increased shear increases the value of $E_{\rm 2\mathrm{H}}$ which is calculated via eq.  \eqref{eq:two_body_energy} by enhancing $v_{\rm rel}^{2}=\|\boldsymbol{v}_1-\boldsymbol{v}_2\|^2$. If we assume the radial density dependence $R^{-0.6}$ of Eq. \eqref{eq:rho_R} used here, and that $v_{\rm rel}^{2}\propto R^{-1}$ then whether BH binaries are more likely to form closer or further from the SMBH depends on the steepness of the exponent of eq. \eqref{eq:cap_criterion} with $\Sigma$ (the surface gensity of the gas, eq. \eqref{eq:rho_R}). For the binary formation probability to be independent of $R$ then it would require the dissipation in the first encounter to scale as $\Delta E_{\rm bin}/E_{\rm \mathrm{H},c}\propto \Sigma^{1.6}$. Here we narrow the dependence down to $\Sigma^{0.90\pm0.44}$. In conjunction, our wind tunnel simulations of \cite{Henry_inprep} narrow this slope down to $1.01\pm0.04$. Thus, as a rough estimate, we expect the probability, $P(R)$, of a highly co-planar BH-BH strong encounter ($r_{\rm min,1} < r_{\mathrm{H}}$) forming a binary successfully to scale as $P(R)\propto R^{-0.6}$. Hence we would expect BH binaries to form more easily closer to the SMBH.

\subsection{Comparison to similar studies}
Our results are a natural extension of the pure 3-body results of \citet{Boekholt_2022} without gas. In \citet{Boekholt_2022}, three-body scatterings of BHs orbiting a SMBH were performed using a purely N-body approach. In the absense of gas, the simulations showed the formation of temporary BH binaries, where binaries with more than 2 encounters are possible for a smaller range of impact parameters, though they are short lived. Where the range of impact parameters that lead to higher encounters numbers becomes exponentially smaller for each additional encounter. Whether they have a deep enough encounter for GWs to induce mergers is dependent on the chaotic evolution of the 3-body system, which can only rarely lead to these chance extremely close encounters. Aside from these extreme cases the lack of a dissipation mechanism leads to their eventual disruption by the SMBH, with probability of surviving multiple $N$ encounters dropping off rapidly with $N$ \citep[e.g][]{Li_and_lai_2022,Boekholt_2022}. As shown in this work, the inclusion of gas leads to the formation of reliably hardened binaries in a single large window in impact parameter space, destroying the fractal structure observed in the paramater space of the impact parameter and number of encounters in \citet{Boekholt_2022}. This results from the inability of the binaries to switch orientations reliably after even one encounter due to the efficiency of the gas-induced energy dissipation enabling the binary to rapidly harden. \citet{DeLaurentiis2022} verify that at exceptionally low ambient gas densities, the fractal structure in the number of encounters vs impact parameter reappears in their 3-body scatterings that include semi-analytic dynamical friction. The recovery of the fractal structure at sufficiently low densities is also hinted in \cite{Henry_inprep}. Where lower density encounters have a greater tendency to flip orientation, most commonly at the boundary between families. Though, the sampling is too course to make any reliable comment on the recovery of the fractal structure.

A closely related work to ours is the gas scattering experiments of \citet{Li_Dempsey_Lai+2022}. In their 2D hydro study of embedded BH scatterings, which uses multiple density values, they show a roughly linearly increasing window in their impact window size as a function of initial ambient density. Instead of varying the radial separation they instead vary the initial azimuthal separation of their satellite BHs, which is possibly why they find nearly all binaries to be formed in a retrograde configuration. In this work we varied the initial radial separation in the AGN disc and scaled the azimuthal separation so the approach time for each simulation was identical. Using this alternate setup, we find both prograde and retrograde binaries, though likewise find the cross section to be larger for retrograde encounters, which make up the RS encounter region, see Sec. \ref{sec:families}. Comparing the outcomes, there is good agreement in the dissipation mechanisms being positively correlated with the local gas density. There is also agreement that for lower AGN densities, the maximum allowed encounter energy is lower, as highlighted in the final figure of our paper, Figure \ref{fig:param_space_analytical_all}. 

Comparing to the recent semi-analytical studies, \citet{DeLaurentiis2022} consider three-body scatterings with an Ostriker dynamical friction prescription to account for gaseous effects. \citet{DeLaurentiis2022} find similar "island" regions of $p$ to our purely gravitational study of \citet{Boekholt_2022}, where binaries may successfully form. Whereas with our inclusion of gas we find only a single central valley of captures. Given they switch off their dynamical friction at the closest approach they neglect the subsequent dissipation between periapsis and their approach back towards the Hill sphere, which can be significant, as shown in Figure \ref{fig:cum_energy_dissipation_fid} for example. This extra dissipation could harden the extremely low angular momentum encounters at the troughs that have the largest apoapses and naturally are more prone to being disrupted by the SMBH. Additionally they find increasing $M_{\rm d}$ leads to the formation window moving to lower $p$ values which is the opposite to our findings. Similarly to \citet{Li_Dempsey_Lai+2022}, this discrepancy in the approach trajectories could be due to their azimuthal separation being far smaller than in this study.

In a sibling paper to this \citep{Henry_inprep}, we investigated the same scenario posed in this paper using a shearing box model using the Eulerian grid code Athena++. The use of the new code allows us to compare the two different numerical approaches to the hydrodynamics and by using a shearing box, maintain a higher resolution inside the Hill sphere for less computational cost. We find excellent agreement on several aspects of the BH-BH encounters. This includes the order of magnitude of dissipation expected for gas assisted encounters, which lies around $10-100$ times the energy of the binary upon reaching a separation of one Hill sphere. Also corroborated is the identification of the three encounter families that lie at low, moderate and high impact parameters, which have the orientations prograde, retrograde and prograde respectively (relative to the orbit about the AGN). The slope of $\Delta E/E_{\rm \mathrm{H},c}$ as a function of $r_{\min,1}/r_{\rm \mathrm{H}}$ is also corroborated to within 10\% percent of the values shown here. A larger sampling of $M_{\rm d}$ is performed in \citep{Henry_inprep} which finds that sufficiently low densities the single formation window splits down the centre into two separate islands of capture. Though the shape of the capture window in \citep{Henry_inprep} is identical to those here (i.e a single window with three types of first encounter), the cross section and location of the window tends to be at lower values of $p$ for the same AGN disc densities. However, comparing the size of the formation window of the global simulations here to the straight wind tunnel of \citep{Henry_inprep} directly is likely not reliable since the linear geometry of the shearing box model will lead to different gravitational focusing during the initial approach. 

\section{Caveats}
\label{sec:caveats}

\begin{itemize}
    \item In this paper we restricted attention to strictly co-planar encounters in the midplane of a 3D gaseous disk, but in reality there is also a 3D distribution of BHs around AGN with some vertical velocity dispersion. Inclination has been shown by \citet{Li_and_lai_2022} to reduce the likelihood of encounter, thus inclination may alter the size of the capture windows shown here. Though they do not include gas in their study. In a follow up to our 2D study in preparation, we extend the previous work in \citet{Boekholt_2022} to 3D.
    \item We assume gas heating/thermal shocks do not significantly increase the gas temperature
    anywhere in the entire simulation domain. While this may be a good assumption for the main body of the annulus distant from the encounter, the high speed intersections of the accretion discs of the BHs will in practice lead to \textit{significant} heating of their material due to their high density and relative velocity. \citet{Tagawa2023_feedback} show that even single black holes can provide a heating mechanism to the local gas through accretion driven jet outflows. Therefore we encourage future studies to include such effects if feasible.
    \item Despite the number of simulations shown here, we are still dealing with far fewer statistics than 3-body studies and recommend further simulations to further probe additional parameters that affect the dissipation processes. For example, can we constrain analytically the slope of the disspation -- encounter depth relation and improve the errors on the gradient overall.
    \item In addition to non inclined orbits, our BHs are initialised with only circular orbits around the SMBH. A more accurate nuclear BH population would include also include a radial velocity dispersion, i.e orbits around the SMBH would have their own eccentricity as well, which would affect the energy of a binary going into an encounter.
    \item We have shown gravitational dissipation and torques from the CSMDs during the encounter can be reproduced by accretion, despite the accretion radius being far larger than its true value by many orders of magnitude. However there is still a factor of a few difference in the strength of these effects between some models. Additionally, as in \citetalias{Rowan2022}, we find the accretion to be super-Eddington by several orders of magnitude. It is possible that a more accurate handling of the accretion leads to results different to both the accretionless and accretion enabled results shown in this paper. However, accretion at such small scales necessitates extreme resolutions and the inclusion of radiative effects which becomes \textit{extremely} computationally expensive.
\end{itemize}

\section{Summary and Conclusions}
\label{sec:conclusions}
Using a hydrodynamic numerical simulations, we performed a quantitative analysis of the binary formation process in AGN discs, with the aim of constraining the likelihood of binary formation depending on encounter parameters and providing the community with physically motivated analytical tools for predicting binary formation. We performed a total of 114 hydrodynamic simulations comprised of three suites, i) a fiducial suite with the expected AGN density for a thin Shakura-Sunyaev disc ii) a suite with ambient density three times higher and iii) a suite identical to the fiducial where accretion is turned off. Captures occur in each of our suites with varying success rates. We summarise our findings into key points below.
\begin{enumerate}
    \item The addition of the gaseous AGN discs leads to captures from a wider range of radial impact parameters $p$ in the disc and leads to consistently close ($\Delta r < 0.1\Hillradius$)  encounters within one large window, unlike the gasless case which has only two very narrow, separate regions. This results from increased gravitational focusing in the lead up to the encounter from the mass of the CSMDs and strong dissipation from gas effects in the immediate lead up to the first periapsis. 
    \item Binary scatterings in higher density environments form more easily in an even larger range of impact parameters. Additionally, they also allow easier formation of binaries that have larger close separations, owed to increased focusing from more massive CSMDs and stronger gaseous energy dissipation.
    \item In our simulations with accretion, dissipation during the encounter is initially dominated by the SMBH whose shear removes energy from the binary. Once the binaries get closer ($\Delta r \lesssim 0.5\Hillradius$) energy is then deposited into the binary from primarily the gas gravity and a little accretion as the CSMDs are perturbed by the opposing BH. During the periapsis passage, when the CSMDs collide, accretion dissipation quickly switches sign and removes energy from the binary very efficiently, leaving it bound.
    \item When accretion is switched off, during the close approach gas can flow past the BHs where they would normally be accreted, leading to a net accumulation of gas behind the BH and inducing a drag. When accretion is enabled, this drag is instead reflected in the accretion dissipation where there is a direct 'headwind' on the BHs. We find little change in the trajectory of the BHs when accretion is disabled and similar energy dissipation, indicating accretion mimics the gas drag well on the small scales around the BHs.

    \item A detailed understanding of the chronology and characteristics of the encounter was developed. Three characteristic orbital trajectories, determined by their impact parameter $p$ were identified (see Figure \ref{fig:enc_families}) which demonstrate specific behaviours in their dissipation processes. In the frame of the inner BH orbiting counterclockwise around the SMBH the trajectory of the three families are as follows:
    \begin{enumerate}
        \item \textit{Leftsided} encounters - encounters with low $p$ that pass to the left of the inner BH which form prograde binaries.
        \item \textit{Rightsided} encounters - encounters with moderate $p$ values that pass to the right of the inner BH and form retrograde binaries
        \item \textit{Turnaround} encounters - encounters with high $p$ that pass initially to the right before looping back on themselves and going round the left of the inner BH, forming prograde binaries.
    \end{enumerate}
    We find that rightsided encounters are most favourable for successful binary formation as they are located in the centre of the space of $p$ for successful binary formation. leftsided and Turnaround encounters impact parameters that place them further from the rightsided undergo encounters with increased minimum separations and are less likely to remain bound due to less accretion and gravitationally induced energy dissipation from the local gas.
    
    \item Strong oscillations in the dissipation from gas gravity, first identified in our last paper \citetalias{Rowan2022}, are again observed in all our simulations where the binaries have deeper encounters due to non-cynlindrically symmetric perturbations to the CSMDs arising from tides induced by the opposing binary BH. These perturbations dominate the gas dissipation during the close approach in both simulations with and without accretion and accretion, though the net dissipation is found to be positive in the former and negative in the latter case. The strength of the oscillations in the binary energy could leave a detectable GW signature, however the orbital frequency of these inhomogeneities is of the order $\sim10^{-7}Hz$ which puts them out of the low frequency sensitivity range of even LISA.
  
    \item We find the dissipation during the first encounter in all simulation suites follows a power-law with the minimum separation. This slope is found to be steeper when the AGN disk density is increased.
    
    \item We construct an analytic criterion for binary formation, eq. \eqref{eq:cap_criterion}. The relation is derived directly from the fully hydrodynamic simulations in this study yet only depends on pre-encounter properties. Thus, it is physically well informed but does not require the user to implement any such physics in order for it to operate, making it well suited to improving population studies such as \cite[e.g.][]{Tagawa2020}. Using the two-body energy at two Hill spheres separation and the impact parameter measured at one Hill sphere separation, the criterion can predict with a success rate >90\% whether the binary will remain bound, for the AGN densities shown in our paper. The accuracy of the predictor is affirmed in Figure \ref{fig:param_space_analytical_all}.

\end{enumerate}

\section*{Acknowledgements}
\begin{itemize}
    \item This work was performed using the Cambridge Service for Data Driven Discovery (CSD3), part of which is operated by the University of Cambridge Research Computing on behalf of the STFC DiRAC HPC Facility (www.dirac.ac.uk). The DiRAC component of CSD3 was funded by BEIS capital funding via STFC capital grants ST/P002307/1 and ST/R002452/1 and STFC operations grant ST/R00689X/1. DiRAC is part of the National e-Infrastructure.
    \item Surface density plots of our simulations were rendered using SPLASH (see \citealt{Price2007}).
    \item  This work was supported by the Science and Technology Facilities Council Grant Number ST/W000903/1. 
    \item This work was supported by NASA ATP grant 80NSSC22K0822 attributed to Zoltan Haiman$^{3,4}$.  
\end{itemize}

%%%%%%%%%%%%%%%%%%%%%%%%%%%%%%%%%%%%%%%%%%%%%%%%%%
\section*{Data Availability}

The data underlying this article will be shared on reasonable request to the corresponding author.
%%%%%%%%%%%%%%%%%%%%%%%%%%%%%%%%%%%%%%%%%%%%%%%%%%

\bibliographystyle{mnras}
\bibliography{Paper}

% do not change these lines
\bsp	% typesetting comment
\label{lastpage}
\end{document}